\documentclass[opre,nonblindrev]{informs3} 

\OneAndAHalfSpacedXI 

\usepackage{natbib}
 \bibpunct[, ]{(}{)}{,}{a}{}{,}%
 \def\bibfont{\small}%
\TheoremsNumberedThrough     
\ECRepeatTheorems

\EquationsNumberedThrough    

\usepackage{thm-restate, enumitem}

\newcommand{\bs}[1]{\boldsymbol{#1}}

\newcommand{\prtwo}[2]{{\mathbb P}_{#1} \left( #2 \right)}

\newcommand{\expartwo}[2]{{\mathbb E}_{#1} [ #2 ]}
\newcommand{\extwo}[2]{{\mathbb E}_{#1} \left[ #2 \right]}
\newcommand{\eps}{\epsilon}
\newcommand{\mylight}{\mathrm{light}}
\newcommand{\myheavy}{\mathrm{heavy}}
\newcommand{\mypower}{\mathrm{Power}}
\newcommand{\mymulti}{\mathrm{multi}}
\newcommand{\mysingle}{\mathrm{single}}
\newcommand{\opt}{\mathrm{OPT}}
\newcommand{\objfunc}{\Psi}
\newcommand{\familyall}{{\cal U}_{\mathrm{all}}}
\newcommand{\familysmall}{{\cal U}_{\mathrm{small}}}
\newcommand{\bbN}{\mathbb{N}}
\newcommand{\bbI}{\mathbb{I}}

\begin{document}


\RUNAUTHOR{Aouad and Segev}

\RUNTITLE{An Approximate Dynamic Programming Approach to The Incremental Knapsack Problem}

\TITLE{An Approximate Dynamic Programming Approach to \\
The Incremental Knapsack Problem}

\ARTICLEAUTHORS{%
\AUTHOR{Ali Aouad}
\AFF{Department of Management Science and Operations, London Business School, Regent's Park, \\ London, United Kingdom NW14SA,  \EMAIL{ aaouad@london.edu}} 
\AUTHOR{Danny Segev}
\AFF{Department of Statistics and Operations Research, School of Mathematical Sciences, \\ Tel Aviv University, Tel Aviv 69978, Israel, \EMAIL{segevdanny@tauex.tau.ac.il}}
} 

\ABSTRACT{%
We study the incremental knapsack problem, where one wishes to sequentially pack items into a knapsack whose capacity expands over a finite planning horizon, with the objective of maximizing time-averaged profits. While various approximation algorithms were developed under mitigating structural assumptions, obtaining non-trivial performance guarantees for this problem in its utmost generality has remained an open question thus far. In this paper, we devise a polynomial-time approximation scheme for general instances of the incremental knapsack problem, which is the strongest guarantee possible given existing hardness results. In contrast to earlier work, our algorithmic approach exploits an approximate dynamic programming formulation. Starting with a simple exponentially sized dynamic program, we prove that an appropriate composition of state pruning ideas yields a polynomially sized state space with negligible loss of optimality. The analysis of this formulation synthesizes various techniques, including new problem decompositions, parsimonious counting arguments, and efficient rounding methods, that may be of broader interest.}%



\maketitle

\section{Introduction}

Integer packing problems have traditionally been some of the most fundamental and well-studied computational questions in discrete optimization. Contributions along this line of research have ranged from advancing the theory of algorithm design to developing practical tools for tackling a wide range of real-life applications. One of the cornerstone resources in this context is the binary knapsack problem. Here, given a finite collection of items, each associated with profit and weight parameters, one wishes to identify the most profitable subset of items subject to a single packing constraint, in the form of a pre-specified upper bound on the total weight of selected items. Despite its seemingly simple formulation, binary knapsack turned out to incorporate just the right amount of combinatorial complexity to end up as one of Karp's 21 NP-complete problems \citeyearpar{karp1972reducibility}, and at the same time, to admit various fully polynomial-time approximation schemes \citep{ibarra1975fast, Lawler79, MagazineO81, KellererP99}. Due to the large body of literature on the binary knapsack problem and its descendants, we refer the avid reader to excellent books dedicated to this topic \citep{TothM90, KellererPP04} for an exhaustive overview.

\paragraph{The incremental knapsack problem.} Knapsack-like problems were studied under various forms and shapes motivated by applications such as financial planning, energy management, and logistics. In particular, the thesis work of \cite{Sharp07} and \cite{Hartline08} introduced a novel multi-period generalization, dubbed the incremental knapsack problem, as part of their incremental optimization modeling framework; this approach extends classic combinatorial optimization problems into multi-period formulations with time-varying constraints. Deferring the finer details of this setting to Section~\ref{subsec:model_description}, rather than having a single knapsack constraint, we are now considering a discrete planning horizon of $T$ periods, which are associated with non-decreasing capacities $W_1 \leq \cdots \leq W_T$ as well as with non-negative coefficients $\lambda_1, \ldots, \lambda_T$. An additional input ingredient is a collection of $n$ items, where each item $i \in [n]$ is characterized by a profit $p_i$ and a weight $w_i$. These items are to be incrementally introduced along a nested sequence of subsets $S_1 \subseteq \cdots \subseteq S_T \subseteq [n]$, subject to time-varying knapsack constraints, stipulating that $w( S_t ) \leq W_t$ for every $t\in [T]$. The goal is to maximize the $\lambda$-averaged profit, $\sum_{t \in [T]} \lambda_t \cdot p ( S_t )$. Here, $w(\cdot)$ and $p(\cdot)$ are extensions of the weight and profit functions from individual items to item sets, meaning that $w( S_t ) = \sum_{i \in S_t} w_i$ and $p( S_t ) = \sum_{i \in S_t} p_i$.

This modeling approach naturally captures resource allocation settings where the available resource (e.g., capital, production capacity, demand) grows in a predictable way over a finite planning horizon. A canonical example of such resource allocation problems is financial planning. For instance, well-informed households generally wish to optimize how their cumulative income is invested over time into low-volatility securities that procure regular payments; from that angle, time-varying discounts reflect specific households' investment profiles. Similarly, this approach can be utilized for modeling the selection of investments in bond markets and various government decision processes, as illustrated by \citet{FaenzaM18}. Additionally, the incremental knapsack problem has certain technical connections with sequential assortment optimization problems in revenue management, as discussed in Section~\ref{subsec:connections}. 

\paragraph{Previous work.} We proceed by demonstrating that this generalization of the standard binary knapsack problem creates intriguing computational challenges. To this end, we briefly summarize known hardness results and algorithmic findings related to the incremental knapsack problem. Further connections to well-studied optimization settings are discussed in Section~\ref{subsec:connections}.
\begin{itemize}
\item {\em Hardness:} Quite surprisingly, \cite{BienstockSY13} and~\cite{Ye16} showed that the incremental knapsack problem is strongly NP-hard. Consequently, the existence of an FPTAS is ruled out in this computational setting unless $\mathrm{P} = \mathrm{NP}$, thus creating a strong separation from its ancestor, the binary knaspack problem, which is known to admit various such approximation schemes \citep{ibarra1975fast, Lawler79, MagazineO81, KellererP99}.

\item {\em Special cases.} Positive results for special cases of the incremental knapsack problem have been appearing in the literature for quite some time. \citet{HartlineS06} devised a polynomial-time $O(1 / \log T)$-approximation for the incremental subset sum problem, where all input parameters are integral and $p_i = w_i$ for every $i\in [n]$. \citet[Chap.~3.3]{Hartline08} subsequently proposed a pseudo-polynomial $O(n W_T)$-time dynamic program that yields a $1/2$-approximation for incremental subset sum. By extending the standard pseudo-polynomial dynamic program for the binary knapsack problem, \citet[Chap.~3.4]{Sharp07} attained an approximation scheme in this setting.  Her algorithm computes a $(1-\eps)$-approximate solution in $O( \frac{1}{\eps}\cdot (nW_T)^{O(T)})$ time, meaning that this approach is exponential in $T$. Subsequently, \citet{CrocePS18} focused on the special case of incremental knapsack with $T=3$ time periods; they developed an LP-based algorithm with a tight approximation ratio of $30/37$.

\item {\em Approximability under parametric assumptions.} More recently, research efforts have been put into algorithmic approaches that are provably good or computationally efficient in specific parametric regimes. For example, by combining LP rounding methods and disjunctive programming ideas, \cite{BienstockSY13} devised a polynomial-time constant-factor approximation; the specific constant attained grows as a function of the minimum parameter $p$ satisfying $W_t/W_1 \leq t^p$ for all $t\in [T]$. Later on, \citet{CrocePS19a} developed an enumeration-based approximation scheme that identifies an $(1-\eps)$-approximate solution in $O((nT)^{O(T/\eps)})$ time; due to its exponential dependency on $T$, this approach corresponds to a PTAS for a fixed number of time periods.

\item {\em Time-invariant incremental knapsack.} The time-invariant formulation constitutes a special case of incremental knapsack for which stronger performance guarantees are known to exist. The latter heavily relies on a simplifying structural assumption, where all time periods are uniformly weighed within the objective function, meaning that $\lambda_1 =  \cdots = \lambda_T$. In this setting, \cite{BienstockSY13} attained a PTAS in the regime where the number of time periods is sublogarithmic in the number of items, i.e., $T = O(\sqrt{\log n})$.  Recently, \citet{FaenzaM18} developed a PTAS for the time-invariant incremental knapsack problem without additional assumptions. Their approach combines techniques such as problem sparsification, compact enumeration, and LP-rounding methods.
\end{itemize}

\paragraph{Research questions.} To summarize, while numerous research papers have shed light on various computational aspects, we are not aware of any way to leverage the above-mentioned advances to obtain non-trivial approximations for the incremental knapsack problem in its utmost generality. Quoting the work of \citet[Sec.~5]{BienstockSY13}: ``{\em Our work leaves the following open questions \ldots Is there a polynomial time algorithm for incremental knapsack with a constant-factor approximation ratio that makes no assumption on the growth rate of $W_t$?}''. Similarly, this question has been restated by \citet[Sec.~3]{FaenzaM18}: ``{\em Following Theorem~1, one could ask for a PTAS for the general incremental knapsack problem}''. Our paper seeks to resolve these fundamental open questions.

\subsection{Our contributions}

The main result of this paper resides in developing a PTAS for the incremental knapsack problem in its utmost generality, i.e., without mitigating assumptions on the underlying number of time periods, item weights, or any other parameter. Namely, we show that the optimal profit can be approached within any fixed degree of accuracy via an algorithmic approach whose running time is polynomial in the input size, as formally stated in Theorem~\ref{thm:general} below; to avoid cumbersome expressions, we make use of $\tilde{O}(\cdot)$ to hide polylogarithmic dependencies, i.e., $\tilde{O}(f(n)) = O( f(n) \cdot ( \log f(n) )^{ O(1) } )$. Recalling that the incremental knapsack problem has previously been shown to be strongly NP-hard \citep{BienstockSY13, Ye16}, this result fully settles its complexity status.

\begin{restatable}{theorem}{mainresult}
\label{thm:general}
For any accuracy level $\eps > 0$, the incremental knapsack problem can be approximated within factor $1-\eps$ of optimal. The running time of our algorithm is $O( n^{ \tilde{O}( 1 / \eps^3 ) } \cdot |{\cal I}|^{ O(1) } )$, where $|{\cal I}|$ stands for the input size in binary representation.
\end{restatable}

From an algorithmic perspective, while existing approximation schemes for special cases of the incremental knapsack problem mainly relied on LP-rounding methods and disjunctive programming \citep{BienstockSY13, FaenzaM18}, our approach is based on approximate dynamic programming. In a nutshell, the foremost ingredient of this formulation is an efficient state pruning method that takes advantage of how certain near-optimal solutions are structured. This approach combines various techniques, 
including newly developed problem decompositions, parsimonious counting arguments, and efficient rounding methods. For ease of exposition, our approximate dynamic programming ideas are presented in an incremental way throughout the technical parts of this paper, as we proceed to explain.

\paragraph{Approximation scheme assuming ${\lambda^+}$-boundedness.} In Section~\ref{sec:alg_bounded}, we begin by devising an approximation scheme whose running time includes a polynomial dependency on $({\sum_{t \in [T]} \lambda_t}) / \lambda_1$. Clearly, the latter ratio by itself could be exponential in the input size, meaning that our proposed algorithm constitutes a PTAS only when $({\sum_{t \in [T]} \lambda_t}) / \lambda_1 = |{\cal I}|^{ O(1) }$, a property referred to as $\lambda^+$-boundedness. To derive this result, we establish a so-called prefix-like property of near-optimal solutions where, by slightly altering item profits, it is optimal to introduce items within each profit class by increasing weights. Given this property, we formulate an intermediate dynamic program whose states encapsulate, among other parameters, a ``count vector'' that specifies the number of items introduced thus far out of certain profit classes. Our main structural finding shows that this exponentially sized state space can be efficiently shrunk with a negligible loss of optimality. Specifically, we describe rounding operations that transform the original count vectors into highly structured vectors, of which there are polynomially many distinct occurrences.

\paragraph{General approximation scheme.} Given these results, Section~\ref{sec:alg_general} is dedicated to devising a PTAS that applies to any instance of the incremental knapsack problem, with no assumptions whatsoever. This general purpose algorithm is developed via an efficient reduction to the $\lambda^+$-bounded setting. To this end, we prove that near-optimal solutions can be decomposed through the connected components of an appropriately defined bipartite graph, representing how items of different profit classes are assigned to ``clusters'' of time periods, each meeting the $\lambda^+$-boundedness property. Moreover, we argue that the best-possible such decomposition can be computed via further dynamic programming ideas.

\subsection{Problem formulation} \label{subsec:model_description}

In an instance of the incremental knapsack problem, we are given a collection of $n$ items, where each item $i \in [n]$ has a profit of $p_i$ and a weight of $w_i$. By reading the remainder of this formulation, it is easy to verify that both parameters can be assumed to be strictly positive. These items may be introduced across a discrete planning horizon that consists of $T$ time periods, indexed $1, \ldots, T$, whose respective knapsack capacities are designated by $W_1 \leq \cdots \leq W_T$. Moreover, each period $t \in [T]$ is associated with a non-negative coefficient $\lambda_t$.

To formally capture how items are inserted over time, we define a solution  ${\cal S} = \langle S_1, \ldots, S_T \rangle$ to be a nested sequence of items sets $S_1 \subseteq \cdots \subseteq S_T$ falling within the above-mentioned capacities, i.e., $w( S_t ) \leq W_t$ for every $t \in [T]$. We refer to the  difference $S_t \setminus S_{t-1}$ as the collection of items introduced in time period $t$, with the convention that $S_0 = \emptyset$, meaning that each set $S_t$ is precisely the collection of items that were introduced up until and including period $t$, with a total profit of $p( S_t )$. The objective function $\objfunc( {\cal S} )$ averages the profits of these item sets over all time periods through their $\lambda$-coefficients, namely, $\objfunc( {\cal S} ) = \sum_{t \in [T]} \lambda_t \cdot p( S_t )$. Our goal is to compute a solution that maximizes this function.

\paragraph{Equivalent representation of ${\objfunc}$.} Moving forward, it will be convenient to rewrite the objective function $\objfunc$ as a summation over ``item contributions'' rather than over time period profits. For this purpose, let $\lambda^{t+} = \sum_{\tau = t}^T \lambda_t$, noting that $\lambda^{1+} \geq \cdots \geq \lambda^{T+}$. We assume without loss of generality that $\lambda^{T+} > 0$, since all time periods $t \in [T]$ with $\lambda_t = 0$ can be eliminated in advance. Additionally, with respect to a solution ${\cal S} = \langle S_1, \ldots, S_T \rangle$ and an item $i \in S_T$, we use $\nu(i,{\cal S})$ to denote the unique time period in which this item was introduced. Consequently, by defining $\objfunc_i ( {\cal S} ) = p_i \cdot \lambda^{ \nu(i,{\cal S}) + }$ for any such item, it follows that the objective function can be decomposed into the sum of individual item contributions as follows:
\begin{equation} \label{eqn:profit_decomp}
\objfunc ( {\cal S} ) ~~=~~ \sum_{t \in [T]} \lambda_t \cdot p( S_t )
~~=~~ \sum_{i \in S_T} \Bigg( p_i \cdot \sum_{ t = \nu(i,{\cal S}) }^T \lambda_t \Bigg) ~~=~~ \sum_{i \in S_T} p_i \cdot \lambda^{ \nu(i,{\cal S}) + } ~~=~~ \sum_{i \in S_T} \objfunc_i ( {\cal S} ) \ .
\end{equation}

\subsection{Connections to additional optimization settings} \label{subsec:connections}

Beyond our preliminary overview of directly related work, it is worth pointing out that the incremental knapsack problem is related to additional streams of literature in discrete optimization. In what follows, we discuss two such connections and highlight the distinctive features of incremental knapsack.

\paragraph{Maximum generalized assignment.} Incremental knapsack has certain similarities with the maximum generalized assignment problem. Here, $n$ items can be assigned to $m$ disjoint knapsacks with capacities $W_1,\ldots, W_m$. Placing item $i \in [n]$ in knapsack $j\in [m]$ consumes $w_{ij}$ units of capacity and generates a profit of $p_{ij}$; we note that both quantities depend on the knapsack to which an item is assigned to. \cite{ShmoysT93} were the first to implicitly study this problem, by presenting an LP-based $2$-approximation for its minimization variant, which provides with small modifications a $1/2$-approximation for the maximization problem, as noted by \cite{ChekuriK05}. The latter paper further identified special cases of generalized assignment that turn out to be APX-hard; the authors also devised a PTAS when each item's profits and weights are uniform across all knapsacks, a setting known as the multiple knapsack problem. Subsequently, \cite{FleischerGMS11} considered the separable assignment problem, which extends generalized assignment, for which they devised an LP-based $(1-1/e)$-approximation. \cite{FeigeV10} broke this barrier by proposing an LP-based algorithm that attains an approximation factor of $1 - 1/e + \delta$, for some absolute constant $\delta > 0$. Additional papers have studied closely related variants, including those of~\cite{CohenKR06} and~\cite{NutovBY06}.

As observed by~\citet{BienstockSY13}, the incremental knapsack problem is reducible to maximum generalized assignment in the special case where the  capacity constraints can be equivalently represented as cardinality constraints on the number of items introduced at each time period. Nevertheless, to our knowledge, connections to this line of research do not lend themselves for the purpose of designing a general-purpose PTAS for the incremental knapsack problem.

\paragraph{Assortment optimization over time.} Interestingly, technical connections can also be established in relation to a class of problems known as assortment optimization over time under the Multinomial Logit (MNL) choice model. In this setting, we are given a collection of products $i\in [n]$ characterized by a profit $p_i$ and a preference weight $w_i$. The firm decides on a sequence of nested subsets of products $S_1\subseteq \cdots \subseteq S_T$ that are examined by customers according to  an exogenous distribution $\lambda_1, \ldots, \lambda_T$. The goal is to maximize the cumulative expected revenue, $\sum_{t \in [T]} \lambda_t \cdot ( \sum_{i \in S_t} p_i \cdot \frac{w_i}{1 + w( S_t ) } )$, where the inner ratio $\frac{w_i}{1 + w( S_t ) }$ reflects the purchase probability of product $i$ within the assortment $S_t$, as prescribed by the MNL model. This problem has recently been introduced by \citet{davis2015assortment}, who proposed a greedy-like $1/2$-approximation when $\lambda_1 = \cdots = \lambda_T$. Subsequently, \citet{GallegoLTW20} developed a dynamic programming-based algorithm that attains an improved approximation ratio of $6 / \pi^2 \approx 0.607$ and actually applies to a more general class of choice models. Focusing on the MNL setting, \citet{aouad2015display} devised a PTAS via an approximate dynamic programming formulation. Closely related variants have been considered in additional papers, including those of~\citet{derakhshan2018product} and~\citet{flores2019assortment}.

To understand the connection between this setting and the incremental knapsack problem, note that the linear-fractional terms $ \sum_{i \in S_t} p_i \cdot \frac{w_i}{1 + w( S_t ) } $ in the objective function indirectly capture soft capacity constraints. Specifically, rather than imposing a hard constraint of the form $w( S_t ) \leq W_t$, the profit contribution of each product in $S_t$ decreases as a function of $w( S_t )$. To formalize this connection, it is not difficult to verify that assortment optimization over time is reducible to incremental knapsack, had we been able to efficiently guess the sequence $w(S^*_1),\ldots,w(S^*_T)$, where $S^*_1,\ldots,S^*_T$ is the optimal sequence of assortments. Unfortunately, we are unaware of any way to reciprocally exploit this connection for the purpose of designing a PTAS for the incremental knapsack problem.

\section{PTAS for $\bs{\lambda^+}$-Bounded Instances} \label{sec:alg_bounded}

In what follows, we devise an approximation scheme for the incremental knapsack problem whose running time incorporates a polynomial dependency on the ratio $\rho = \lambda^{1+} / \lambda^{T+}$ between the largest and smallest $\lambda^+$ values. The specifics of this result are formally stated in the next theorem.

\begin{theorem}\label{thm:main_bounded_lambda-0}
For any accuracy level $\eps > 0$, the incremental knapsack problem can be approximated within factor $1-\eps$ of optimal in $O((n \rho)^{ \tilde{O}( 1 / \eps^2 ) } \cdot |{\cal I}|^{ O(1) } )$ time.
\end{theorem}

\subsection{The inverse problem}

\paragraph{Auxiliary formulation.} To provide an accessible account of this result, we introduce a closely related variant, dubbed inverse incremental knapsack. Here, given a profit requirement $\phi\geq 0$, suppose that our goal is to identify a solution ${\cal S} = \langle S_1, \ldots, S_T \rangle$ that minimizes the total weight of inserted items, $w(S_T)$, subject to obtaining a profit of at least $\phi$, namely, $\Psi({\cal S}) \geq \phi$. Moving forward, we use  ${\cal S}^{*\phi}$ to designate a fixed optimal solution for the inverse problem, parameterized by $\phi$; when the latter problem is infeasible, we indicate this situation by setting ${\cal S}^{*\phi} = \bot$. In the remainder of this section, we develop an approximation scheme for computing a super-optimal solution to the inverse incremental knapsack problem, at the expense of slightly violating the profit constraint. As formally stated in the next claim, our algorithm will incur a polynomial dependency on $\rho$ as part of its running time.

\begin{theorem} \label{thm:main_bounded_lambda}
For any accuracy level $\eps > 0$, given a profit requirement $\phi \geq 0$ for which ${\cal S}^{*\phi}  \neq \bot$, we can compute in $O((n \rho)^{ \tilde{O}( 1 / \eps^2 ) } \cdot |{\cal I}|^{ O(1) } )$ time a solution ${\cal S}$ satisfying:
\begin{enumerate}[label=(\alph*)]
\item Super-optimality: $w(S_T) \leq w( S^{*\phi}_T )$.

\item $\eps$-profit violation: $\Psi({\cal S})\geq (1-\eps)\cdot \phi$.
\end{enumerate}
\end{theorem}

\paragraph{Immediate consequence.} To derive Theorem~\ref{thm:main_bounded_lambda-0}, the important observation is that our algorithm for the inverse formulation can be leveraged to devise an approximation scheme for incremental knapsack in a straightforward way. To this end, letting ${\cal S}^*$ be an optimal solution for the latter problem, suppose we were given in advance an estimate $\phi \in [(1 - \eps) \cdot \Psi({\cal S}^*), \Psi({\cal S}^*)]$ for the optimal profit. It is easy to verify that, since $\Psi({\cal S}^*) \in [ \lambda_{\min} p_{\min}, n \lambda_{\max} p_{\max} ] \cup \{ 0 \}$, such an estimate can be obtained by enumerating over either $0$ or quantities of the form $\lambda_{\min} p_{\min} \cdot (1+\eps)^k$ within this interval, for integer values of $k$. Consequently, whenever $\Psi({\cal S}^*) \geq \phi$, the $\phi$-parameterized inverse formulation is clearly feasible, meaning that by applying Theorem~\ref{thm:main_bounded_lambda}, we identify a solution ${\cal S}$ satisfying $\Psi({\cal S})\geq (1-\eps)\cdot \phi\geq (1-2\eps)\cdot \Psi({\cal S}^*)$, which is a $(1-2\eps)$-approximate solution to the original incremental knapsack problem. As a side note, it is worth pointing out that the latter argument does exploit the super-optimality of ${\cal S}$; this property will play a crucial role in Section~\ref{sec:alg_general}, once we consider arbitrarily structured instances.

\subsection{Preliminaries: Prefix-like near-optimal solutions}

Focusing on an inverse incremental knapsack instance parameterized by $\phi$, it would be convenient to fix the profit requirement $\phi\geq0$ from this point on. For simplicity of notation, we further drop the symbol $\phi$ when referring to ${\cal S}^{*\phi}$, which results in having ${\cal S}^*$ as an optimal solution to the inverse problem instance being considered.

\paragraph{Profit classes.} We begin by scaling all profits such that $\min_{i \in [n]} p_i = 1$. In addition, we round down the profit of each item to the nearest integer power of $1+\eps$. It is easy to verify that, as a result of this alteration, we are losing at most an $\eps$-fraction of the profit $\Psi( {\cal S}^* )$; clearly, once the original item profits are restored, the profit of any solution can only increase. For simplicity, we keep denoting the modified profits by $p_1, \ldots, p_n$. Following these preprocessing steps, for every $\ell \in \bbN_0$, we use ${\cal P}_{\ell} = \{ i \in [n] : p_i = (1+\eps)^{\ell} \}$ to designate the collection of items associated with a profit of $(1+\eps)^{\ell}$; this set will be referred to as the profit class ${\cal P}_{\ell}$.

\paragraph{Prefix-like solutions.} For any profit class ${\cal P}_{ \ell }$ and for any integer $k \in [0, |{\cal P}_{ \ell }|]$, we use ${\cal P}_{ \ell }[k]$ to denote the set of $k$ smallest weight items in ${\cal P}_{ \ell }$, breaking ties arbitrarily. Also, ${\cal P}_{ \ell }[k_1,k_2]$ will be the set of $k_1$-th smallest, $(k_1 + 1)$-th smallest, up to $k_2$-th smallest weight items in ${\cal P}_{ \ell }$, that is, ${\cal P}_{ \ell }[k_1,k_2] = {\cal P}_{ \ell }[k_2] \setminus {\cal P}_{ \ell }[k_1-1]$. We say that a solution ${\cal S} = \langle S_1, \ldots, S_T \rangle$ has a prefix-like structure when, for every time period $t \in [T]$ and class ${\cal P}_{ \ell }$, the items picked by $S_t$ from ${\cal P}_{ \ell }$ are those with smallest weight, i.e., $S_t \cap {\cal P}_{ \ell } = {\cal P}_{ \ell }[ | S_t \cap {\cal P}_{ \ell } | ]$.

Crucial to our algorithmic approach is the next lemma, which argues that the sequence of profit classes ${\cal P}_0, {\cal P}_1, \ldots$ contains an $O( \frac{ 1 }{ \eps } \log \frac{ n \rho }{ \eps } )$-length interval such that the union of its classes has a near-optimal prefix-like solution.

\begin{lemma} \label{lem:structure_sol}
There exist an $O( \frac{ 1 }{ \eps } \log \frac{ n \rho }{ \eps } )$-length interval ${\cal L} \subseteq \bbN_0$ and a prefix-like solution $\tilde{\cal S} = \langle \tilde{S}_1, \ldots, \tilde{S}_T \rangle$ such that:
\begin{enumerate}[label=(\alph*)]
\item The solution $\tilde{\cal S}$ only introduces items from profit classes $\{ {\cal P}_{\ell} \}_{\ell \in {\cal L}}$. \label{item:structure_sol_classes}

\item Super-optimality: $w( \tilde{S}_T ) \leq w( {S}^*_T )$. \label{item:structure_sol_weight}

\item $\eps$-profit violation: $\objfunc( \tilde{\cal S} ) \geq (1 - \eps) \cdot \objfunc( {\cal S}^* )$. \label{item:structure_sol_profit}
\end{enumerate}
\end{lemma}
\proof{\em Proof.}
To determine the interval ${\cal L}$, let $\ell_{\max}$ be the maximal index of a profit class of which ${\cal S}^*$ introduces at least one item, namely, $\ell_{\max} = \max \{ \ell \in \bbN_0 : S_T^* \cap {\cal P}_{\ell} \neq \emptyset \}$. Our interval is defined as ${\cal L} = [ [\ell_{\max} - L + 1]^+, \ell_{\max}]$, where $L = \lceil \log_{1 + \eps} ( \frac{ n \rho }{ \eps } ) \rceil$, meaning that it is indeed of the required length. Having specified the interval ${\cal L}$, we construct a corresponding prefix-like solution $\tilde{\cal S} = \langle \tilde{S}_1, \ldots, \tilde{S}_T \rangle$ by setting $\tilde{S}_t = \bigcup_{\ell \in {\cal L}} {\cal P}_{ \ell }[ | S_t^* \cap {\cal P}_{ \ell } | ]$ for every $t \in [T]$. In other words, for every $\ell \in {\cal L}$, the set $\tilde{S}_t$ picks out of the profit class ${\cal P}_{ \ell }$ exactly the same number of items as $S_t^*$ does; however, these items are picked by order of increasing weight.

One straightforward observation is that, since no items are introduced out of profit classes $\{ {\cal P}_{\ell} \}_{\ell \notin {\cal L}}$, property~\ref{item:structure_sol_classes} is clearly satisfied. In addition, this definition guarantees that $\tilde{\cal S}$ is indeed a feasible solution by noting that, for every time period $t \in [T]$,
\[ w ( \tilde{S}_t ) ~~=~~ \sum_{\ell \in {\cal L}} w \left( {\cal P}_{ \ell } \left[ \left| S_t^* \cap {\cal P}_{ \ell } \right| \right] \right) ~~\leq~~ \sum_{\ell \in {\cal L}} w \left( S_t^* \cap {\cal P}_{ \ell } \right) ~~\leq~~ \sum_{\ell \in \bbN_0} w \left( S_t^* \cap {\cal P}_{ \ell } \right) ~~=~~ w \left( S^*_t \right) \ , \]
where the first inequality is derived by observing that $S_t^* \cap {\cal P}_{ \ell }$ is a subset of $|S_t^* \cap {\cal P}_{ \ell }|$ items in profit class ${\cal P}_{\ell}$, while ${\cal P}_{ \ell } [ | S_t^* \cap {\cal P}_{ \ell } | ]$ is the lightest subset of ${\cal P}_{\ell}$ consisting of that many items. Beyond feasibility, the inequality we have just proven also shows that property~\ref{item:structure_sol_weight} is satisfied.

To prove property~\ref{item:structure_sol_profit}, we first argue that for every $\ell \in {\cal L}$,
\begin{eqnarray*}
\sum_{i \in \tilde{S}_T \cap {\cal P}_{\ell}} \objfunc_i ( \tilde{\cal S} ) & = & \sum_{t \in [T]} \sum_{i \in (\tilde{S}_t \setminus \tilde{S}_{t-1}) \cap {\cal P}_{\ell}} \objfunc_i ( \tilde{\cal S} ) \\
& = & (1 + \eps)^{ \ell } \cdot \sum_{t \in [T]} \lambda^{t+} \cdot \left| (\tilde{S}_t \setminus \tilde{S}_{t-1}) \cap {\cal P}_{\ell} \right| \\
& = & (1 + \eps)^{ \ell } \cdot \sum_{t \in [T]} \lambda^{t+} \cdot \left| (S_t^* \setminus S_{t-1}^*) \cap {\cal P}_{\ell} \right| \\
& = & \sum_{t \in [T]} \sum_{i \in (S_t^* \setminus S_{t-1}^*) \cap {\cal P}_{\ell}} \objfunc_i \left( {\cal S}^* \right) \\
& = & \sum_{i \in S_T^* \cap {\cal P}_{\ell}} \objfunc_i \left( {\cal S}^* \right) \ .
\end{eqnarray*}
Here, the second equality holds since $\objfunc_i ( \tilde{\cal S} ) = p_i \cdot \lambda^{ \nu(i,\tilde{\cal S}) + } = (1 + \eps)^{ \ell } \cdot \lambda^{t+}$ for all items $i \in (\tilde{S}_t \setminus \tilde{S}_{t-1}) \cap {\cal P}_{\ell}$, whereas the fourth equality follows from an analogous argument with respect to ${\cal S}^*$. As a result, based on representation~\eqref{eqn:profit_decomp} of the objective function, to show that $\objfunc( \tilde{\cal S} ) \geq (1 - \eps) \cdot \objfunc( {\cal S}^* )$, it suffices to argue that $\sum_{i \in S_T^* \setminus \tilde{S}_T} \objfunc_i( {\cal S}^* ) \leq \eps \cdot \objfunc( {\cal S}^* )$. For this purpose, note that
\begin{eqnarray*}
\sum_{i \in S_T^* \setminus \tilde{S}_T} \objfunc_i \left( {\cal S}^* \right) & = & \sum_{i \in S_T^* \setminus \tilde{S}_T} p_i \cdot \lambda^{ \nu(i,{\cal S}^*) + } \\
& \leq & \left| S_T^* \setminus \tilde{S}_T \right| \cdot \lambda^{ 1+ } \cdot (1 + \eps)^{\ell_{\max} - L} \\
& \leq & n \cdot \lambda^{ 1+ } \cdot (1 + \eps)^{\ell_{\max}} \cdot \frac{ \eps }{ n \rho } \\
& = & \eps \cdot \lambda^{ T+ } \cdot (1 + \eps)^{\ell_{\max}} \\
& \leq & \eps \cdot \objfunc \left( {\cal S}^* \right) \ .
\end{eqnarray*}
Here, the first inequality holds since $\lambda^{1+} \geq \cdots \geq \lambda^{T+}$ and since $S_T^* \setminus \tilde{S}_T$ is necessarily a subset of $\bigcup_{\ell \in [0,[\ell_{\max} - L]^+]} {\cal P}_{\ell}$, by definition of $\ell_{\max}$, meaning that every item $i \in S_T^* \setminus \tilde{S}_T$ has a profit of at most $(1 + \eps)^{\ell_{\max} - L}$. The second inequality is obtained by observing that $(1 + \eps)^{-L} \leq \frac{ \eps }{ n \rho }$ in view of our choice for $L = \lceil \log_{1 + \eps} ( \frac{ n \rho }{ \eps } ) \rceil$. The last inequality holds since, by definition of $\ell_{\max}$, at least one item $i \in {\cal P}_{ \ell_{\max} }$ is introduced by ${\cal S}^*$, implying that its individual profit contribution $\objfunc_i ( {\cal S}^* )$ is at least $\lambda^{ T+ } \cdot (1 + \eps)^{\ell_{\max}}$. \halmos
\endproof

\subsection{Dynamic programming formulation} \label{subsec:exact_dp}

Taking advantage of Lemma~\ref{lem:structure_sol}, we restrict our attention to prefix-like solutions with respect to the profit classes $\{ {\cal P}_{ \ell } \}_{ \ell \in {\cal L} }$. For convenience, rather that representing such solutions as a sequence of items sets ${\cal S} = \langle S_1, \ldots, S_T \rangle$, we use instead an equivalent sequence of utilization vectors ${\cal N} = \langle N_1, \ldots, N_T \rangle$. Here, each $N_t$ is an integer-valued vector of dimension $|{\cal L}|$, in which each coordinate $N_{t,\ell}$ specifies the number of items picked thus far from the profit class ${\cal P}_{ \ell }$, i.e., $N_{t,\ell} = |S_t \cap {\cal P}_{\ell}|$.


\paragraph{States.} Each state of our dynamic program is described by two parameters:
\begin{itemize}
\item The index of the current period, $t$. This parameter takes one of the values $0, 1, \ldots, T$.

\item The collection of items utilized thus far, specified by the utilization vector $N_t$. This parameter belongs to the family of vectors $\familyall = [|{\cal P}_{\min {\cal L}}|]_0 \times \cdots \times [|{\cal P}_{\max {\cal L}}|]_0$.
\end{itemize}
It is important to emphasize that the family of vectors we are currently working with is of cardinality $| \familyall | = \Omega( n^{ \Omega(|{\cal L}|) } ) = \Omega( n^{ \Omega( \frac{ 1 }{ \eps } \log \frac{ n \rho }{ \eps } ) } ) $, meaning that the state space described so far is exponentially sized. Our goal in subsequent sections will be to develop a more compact, and yet still near-optimal state space representation.

\paragraph{Value function.} The value function $F(t, N_t)$ stands for the maximum-possible accumulated profit contribution $\objfunc ( N_{[1,t]} ) = \sum_{\tau \in [t]} ( \lambda^{\tau+} \cdot \sum_{\ell \in {\cal L}} (1 + \eps)^{ \ell } \cdot ( N_{\tau,\ell} - N_{\tau-1,\ell} ) )$ of a $t$-period solution $N_{[1,t]} = \langle N_1, \ldots, N_t \rangle$ that satisfies the following conditions:
\begin{enumerate}
\item {\em Utilization vector $N_t$ at period $t$}: The set of items introduced across time periods $1, \ldots, t$ is  described by the utilization vector $N_t$.

\item {\em Feasibility}: All capacities are satisfied, i.e., $w( N_{ \tau } ) \leq W_{ \tau }$ for every $\tau \in [t]$, with the convention that $w( N_{ \tau } ) = \sum_{\ell \in {\cal L}} w( {\cal P}_{ \ell }[N_{\tau,\ell}] )$.
\end{enumerate}
When there is no solution satisfying conditions~1 and~2, we define $F(t, N_t) = -\infty$. Consequently, we are interested in finding a utilization vector $N_T \in \familyall$ that minimizes $w(N_T)$ out of those that meet the criterion $F(T,N_T) \geq (1-\eps) \cdot\phi$. At this stage, it is worth pointing out that, by considering the prefix-like solution $\tilde{\cal S}$ constructed in Lemma~1, it follows that the utilization vector $\tilde{N}_T$ associated with $\tilde{S}_T$ satisfies $F(T,\tilde{N}_T) \geq (1 - \eps) \cdot \objfunc( {\cal S}^* ) \geq (1-\eps) \cdot \phi$ and $w( \tilde{N}_T ) \leq w( {S}^*_T )$. Therefore, the previously mentioned criterion is met by some utilization vector, and concurrently, since we return the one whose total weight is minimized, the items introduced have a total weight of at most $w( {S}^*_T )$. Hence, by following the actions taken by our dynamic program, we obtain an inverse incremental knapsack solution that satisfies the properties required in Theorem~\ref{thm:main_bounded_lambda}.

\paragraph{Recursive equations.} Next, we argue that the value function $F$ satisfies a collection of recursive equations, leading to a dynamic programming formulation of the inverse incremental knapsack problem. The important observation is that the feasibility of any $t$-period solution $N_{[1,t]} = \langle N_1, \ldots, N_t \rangle$ is equivalent to jointly ensuring that:
\begin{itemize}
\item The induced $(t-1)$-period solution $N_{[1,t-1]}$ is feasible by itself.

\item Our extension to period $t$ preserves feasibility, meaning that $N_{t-1} \leq N_t$ and $w(N_t) \leq W_t$.
\end{itemize}
Consequently, for every $t\in [T]$ and $N_t\in \familyall$, the value function $F$ can be recursively expressed as
\begin{eqnarray*}
F\left(t,N_t\right) = & \max\limits_{N_{t-1} \in \familyall} & \left\{ F\left(t-1,N_{t-1}\right) + \lambda^{ t+ } \cdot \sum_{\ell \in {\cal L}} (1 + \eps)^{ \ell } \cdot ( N_{t,\ell} - N_{t-1,\ell} ) \right\} \\
& {\rm s.t.} & N_{t-1} \leq N_t \ , \\
& & w(N_t) \leq W_t \ .
\end{eqnarray*}
To better understand the recursion above, note that the expression $\lambda^{ t+ } \cdot \sum_{\ell \in {\cal L}} (1 + \eps)^{ \ell } \cdot ( N_{t,\ell} - N_{t-1,\ell} )$ is precisely the marginal profit contribution of the items introduced in period $t$, bearing in mind that we have already accumulated profit contributions amounting to $F(t-1,N_{t-1})$ up to and including period $t-1$. Specifically, on top of the latter quantity, in period $t$ we are introducing $N_{t,\ell} - N_{t-1,\ell}$ items from each profit class ${\cal P}_{ \ell }$, whose profits are all uniformly equal to $(1 + \eps)^{ \ell }$. When there is no extendible $(t-1)$-period solution, i.e., one satisfying $N_{t-1} \leq N_t$ and $w(N_t) \leq W_t$, we have $F(t,N_{t}) = -\infty$. Finally, the terminal states of this dynamic program are reached at $t=0$, in which case $F(0, 0) = 0$ and $F(0, N_0) = -\infty $ for all other states.

\subsection{Approximate dynamic program: High-level overview} \label{subsec:overview}

\paragraph{Restriction to ${\familysmall}$.} In what follows, we explain how our original dynamic programming formulation can be altered, when considering an appropriately restricted subset of the utilization vectors $\familyall$. To this end, let $\familysmall \subseteq \familyall$ be any family of utilization vectors with $0 \in \familysmall$. Given such a set, one can define an approximate dynamic program $\tilde{F}_{\familysmall}$ that would mimic the exact program $F$, when restricted to the family $\familysmall$. That is, while each state is still of the form $(t, N_t)$, we now require that $N_t \in \familysmall$ instead of $N_t \in \familyall$. In addition, rather than searching over $N_{t-1} \in \familyall$, we substitute this constraint with $N_{t-1} \in \familysmall$. As a result, the recursive equations we obtain for $\tilde{F}_{\familysmall}(t, N_t) $ are given by:
\begin{eqnarray}
\tilde{F}_{\familysmall}\left(t,N_t\right) = & \max\limits_{N_{t-1} \in \familysmall} & \left\{ \tilde{F}_{\familysmall}\left(t-1,N_{t-1}\right) + \lambda^{ t+ } \cdot \sum_{\ell \in {\cal L}} (1 + \eps)^{ \ell } \cdot ( N_{t,\ell} - N_{t-1,\ell} ) \right\} \label{eq:ADP}\\
& {\rm s.t.} & N_{t-1} \leq N_t \ , \nonumber \\
& & w(N_t) \leq W_t \ . \nonumber
\end{eqnarray}
It is easy to verify that this dynamic program is well-defined for any possible choice of $\familysmall \subseteq \familyall$ with $0 \in \familysmall$. However, there is clearly a loss in optimality resulting from the elimination of $\familyall \setminus \familysmall$, corresponding to further restrictions on the utilization vectors being considered. That said, every utilization vector $N_T \in \familysmall$ with $\tilde{F}_{\familysmall}(T,N_T) \neq -\infty$ can still be converted in a straightforward way into a solution ${\cal S}$ with a profit of $\Psi({\cal S}) = \tilde{F}_{\familysmall}(T,N_T)$ and a total weight of $w(S_T) = w(N_T)$.

\paragraph{The fundamental question.} Let $\opt$ be the maximum value of $F(T,N_T)$ over all utilization vectors $N_T \in \familyall$ with $w(N_T) \leq w( S^*_T )$; we remind the reader that ${\cal S}^*$ is a fixed optimal solution to the inverse incremental knapsack instance being considered. As noted in Section~\ref{subsec:exact_dp}, the utilization vector $\tilde{N}_T$, corresponding to the prefix-like solution $\tilde{\cal S}$ given by Lemma~\ref{lem:structure_sol}, satisfies $F(T,\tilde{N}_T) \geq (1-\eps) \cdot\phi$ and  $w(\tilde{N}_T) \leq w( S^*_T )$, immediately showing that $\opt \geq (1-\eps) \cdot\phi$.  Similarly, let us define $\opt_{\familysmall}$ as the maximum value of $\tilde{F}_{\familysmall}(T, N_T)$ over all utilization vectors $N_T \in \familysmall$ with $w(N_T) \leq w( S^*_T )$. The intrinsic question is: {\em Can we efficiently construct a small-sized family $\familysmall \subseteq \familyall$ of utilization vectors for which $\opt_{\familysmall} = (1-O(\eps))\cdot\opt$?}

In the remainder of this section, we will present an explicit construction of a family $\familysmall$ formed by $O ( (n \rho)^{ \tilde{O}( 1 / \eps^2 ) } \cdot \log ( \frac{ w_{\max} }{ w_{\min} } ) )$ utilization vectors,  for which $\opt_{\familysmall} \geq (1-2\eps)\cdot\opt \geq (1-3\eps) \cdot \phi$. Consequently, our algorithm returns the solution ${\cal S}$ associated with the utilization vector $N_T \in \familysmall$ that minimizes $w(N_T)$ over those meeting the criterion $\tilde{F}_{\familysmall}(T, N_T) \geq (1-3\eps)\cdot \phi$. By the preceding discussion, this solution is guaranteed to simultaneously satisfy $w(S_T) \leq w( S^{*}_T )$ and $\Psi({S})\geq (1-3\eps)\cdot \phi$, thereby completing the proof of  Theorem~\ref{thm:main_bounded_lambda}.

\paragraph{Construction of ${\familysmall}$.} In the upcoming discussion, we will constructively define the family of utilization vectors $\familysmall$ in an incremental way. Given an arbitrary utilization vector $N \in \familyall$, we describe two alterations for transforming the latter into a highly structured vector, $N^{(2)}$. Informally speaking, this vector preserves the main features of $N$, in the sense of generating a nearly-matching total profit while still being feasible with respect to the time period capacities. However, the additional structural properties entailed by our alterations would ensure that the family of vectors $\familysmall = \{ N^{(2)} : N \in \familyall \}$ is polynomially sized. Specifically, in Section~\ref{subsec:up_rounding}, we begin by defining the up-rounding operation, where $N$ is mapped to a vector $N^{(1)}$ by appropriately rounding up some of its coordinates. It is important to emphasize that up-rounding by itself may very well lead to capacity violations at various time periods. Next, in Section~\ref{subsec:truncation}, we define the truncation operation, where $N^{(1)}$ is mapped in turn to the final utilization vector $N^{(2)}$. While its finer details are somewhat involved, at least intuitively, this operation discards certain  items within each profit class in order to offset the excess capacity within the up-rounding $N^{(1)}$, and thus, to restore feasibility.

It is important to point out that the upcoming description of our up-rounding and truncation operations is primarily meant for purposes of analysis. Algorithmically speaking, the composition of these operations is not an efficient method for generating the reduced family of utilization vectors ${\familysmall}$, and its naive application would require an explicit enumeration over $\familyall$. That said, the mappings $N \mapsto N^{(1)} \mapsto N^{(2)}$ will be instrumental in bounding the optimality loss entailed by the approximate formulation $\tilde{F}_{\familysmall}$ relative to the exact program $F$. The computational aspects of our approximate dynamic program, including an efficient construction of ${\familysmall}$, are separately discussed in Section~\ref{subsec:analysis}.

\subsection{The up-rounding operation: Definitions and properties} \label{subsec:up_rounding}

In what follows, we define an explicit mapping from each utilization vector $N \in \familyall$ to its up-rounding $N^{(1)}$, which is a new utilization vector with very specific structure.

\paragraph{Light and heavy classes.} To avoid cumbersome notation, we assume without loss of generality that $1 / \eps$ is an integer and that $\eps < 1/4$. We begin by defining a vector-specific dichotomy of the profit classes $\{ {\cal P}_{ \ell } \}_{\ell \in {\cal L}}$, depending on the number of items picked by the utilization vector $N$ out of each one. For this purpose, the profit class ${\cal P}_{ \ell }$ is said to be light for $N$ when $N_{ \ell } \leq 1 / \eps$. The collection of all light classes is denoted by $\mylight( N )$. Every other class, with strictly more than $1 / \eps$ items, is called heavy for $N$, with the analogous notation $\myheavy( N )$.

\paragraph{Constructing ${N^{(1)}}$ via up-rounding.} Each coordinate of the vector $N^{(1)}$ will be defined depending on whether its corresponding profit class is light or heavy. Specifically, out of each light class, $N^{(1)}$ picks exactly the same number of items as $N$ does, meaning that $N^{(1)}_{ \ell } = N_{ \ell }$ for every $\ell \in \mylight( N )$. In contrast, for each heavy class $\ell \in \myheavy( N )$, the way we define $N^{(1)}_{ \ell }$ is more elaborate:
\begin{itemize}
\item We begin by introducing a shorthand notation $W_{ \myheavy }( N )$ for the total weight of all items picked by $N$ out of heavy classes, beyond the $1 / \eps$ lightest ones in each class, i.e.,
    \[ W_{ \myheavy }( N ) = \sum_{ \ell \in \myheavy( N ) } w \left( {\cal P}_{ \ell } \left[ \frac{ 1 }{ \eps } + 1, N_{\ell} \right] \right) \ . \]

\item For every $\ell \in \myheavy( N )$, let ${\cal E}_{ \ell }$ be an over-estimate for ${\cal P}_{ \ell }$'s contribution $w ( {\cal P}_{ \ell } [ \frac{ 1 }{ \eps } + 1, N_{\ell} ] )$ toward the sum $W_{ \myheavy }( N )$, up to an additive factor of $\mypower_2 [ \frac{ \eps }{ |{\cal L}| } \cdot W_{ \myheavy }( N ) ]$. Here, the operator $\mypower_2 [ \cdot ]$ rounds its non-negative argument up to the nearest integer power of $2$, with the convention that $\mypower_2 [ 0 ] = 0$. Specifically, our estimate is of the form ${\cal E}_{ \ell } = \mu_{ \ell } \cdot \mypower_2 [ \frac{ \eps }{ |{\cal L}| } \cdot W_{ \myheavy }( N ) ]$, where $\mu_{ \ell } \geq 1$ is the unique integer for which
    \begin{equation} \label{eqn:def_mu}
    \left( \mu_{ \ell } - 1 \right) \cdot \mypower_2 \left[ \frac{ \eps }{ |{\cal L}| } \cdot W_{ \myheavy }( N ) \right] ~~<~~ w \left( {\cal P}_{ \ell } \left[ \frac{ 1 }{ \eps } + 1, N_{\ell} \right] \right) ~~\leq~~ \mu_{ \ell } \cdot \mypower_2 \left[ \frac{ \eps }{ |{\cal L}| } \cdot W_{ \myheavy }( N ) \right] \ .
    \end{equation}

\item Finally, out of the profit class ${\cal P}_{ \ell }$ in order of increasing weights, the vector $N^{(1)}$ picks the maximum number of items beyond the first $1 / \eps$ ones such that their total weight is still within the estimate ${\cal E}_{ \ell }$. That is,
    \begin{eqnarray} \label{eq:def:N1} 
    N^{(1)}_{ \ell } = \max \left\{ k \in \left[ \frac{ 1 }{ \eps } + 1, | {\cal P}_{ \ell } | \right] : w \left( {\cal P}_{ \ell } \left[ \frac{ 1 }{ \eps } + 1, k \right] \right) \leq {\cal E}_{ \ell } \right\} \ . 
    \end{eqnarray}
\end{itemize}

\paragraph{Is ``up-rounding'' really rounding up?} Based on the above definitions, it is not entirely clear whether we are indeed rounding up each coordinate of $N$ to obtain $N^{(1)}$, or whether some coordinates can actually decrease. To show that $N \leq N^{(1)}$, we first observe that for $\ell \in \mylight(N)$, one clearly has $N^{(1)}_{ \ell } = N_{ \ell }$ by definition. For $\ell \in \myheavy(N)$, due to our method for choosing $N^{(1)}_{ \ell }$ in~\eqref{eq:def:N1}, we necessarily have $N^{(1)}_{ \ell } \geq N_{\ell}$, since $w ( {\cal P}_{ \ell } [ \frac{ 1 }{ \eps } + 1, N_{\ell} ] ) \leq {\cal E}_{ \ell }$ by definition of ${\cal E}_{ \ell }$. These observations also imply that $\mylight( N^{(1)} ) = \mylight( N )$ and $\myheavy( N^{(1)} ) = \myheavy( N )$, meaning that the up-rounding operation preserves the light and heavy classes.

\paragraph{Key properties.} The next lemma summarizes the main properties of the up-rounding operation that will be utilized in our subsequent analysis.

\begin{lemma} \label{lem:properties_up_rounding}
The up-rounding operation satisfies the following properties:
\begin{enumerate}[label=(\alph*)]
\item Monotonicity: $N^{(1)} \leq N^{+(1)}$ for every $N \leq N^+$. \label{lem:rounding_monotone}

\item Weight beyond first $1/\eps$ items: $W_{ \myheavy }( N^{(1)} ) \leq (1 + 2\eps) \cdot W_{ \myheavy }( N )$. \label{lem:rounding_weight}
\end{enumerate}
\end{lemma}

For ease of presentation, since the proof is rather technical, its specifics are deferred to Appendix~\ref{subsec:proof_lem_properties_up_rounding}. At a high level, the above-mentioned properties stem from appropriately choosing the multiplier $\mypower_2 [ \frac{ \eps }{ |{\cal L}| } \cdot W_{ \myheavy }( N ) ]$. By selecting a sufficiently small multiplier, that explicitly depends on the total weight quantity $W_{ \myheavy }( N )$ and the number of classes $|{\cal L}|$, we ensure that the excessive items due to our up-rounding operation only account for an $O(\eps)$ fraction of the total weight, as stated in property~\ref{lem:rounding_weight}. In contrast, arguing for the monotonicity of this operation, corresponding to property~\ref{lem:rounding_monotone}, is more challenging. Here, choosing the multiplier $\mypower_2 [ \frac{ \eps }{ |{\cal L}| } \cdot W_{ \myheavy }( N ) ]$ as an integer power of $2$ will play an important role, as further explained in Appendix~\ref{subsec:proof_lem_properties_up_rounding}.

\subsection{The truncation operation: Definitions and properties} \label{subsec:truncation}

In what follows, given a utilization vector $N$ and its up-rounding $N^{(1)}$, we explain how to define the truncation $N^{(2)}$ of the latter vector. Prior to presenting this construction, we remind the reader that the up-rounding operation was shown in Section~\ref{subsec:up_rounding} to be class-preserving, meaning that $\mylight( N^{(1)} ) = \mylight( N )$ and $\myheavy( N^{(1)} ) = \myheavy( N )$.

\paragraph{Constructing ${N^{(2)}}$ via truncation.} Similarly to the up-rounding operation, each coordinate of the vector $N^{(2)}$ will be defined depending on whether its corresponding profit class is light or heavy. Specifically, out of each light class, $N^{(2)}$ picks exactly the same number of items as $N^{(1)}$ and $N$ do, that is, $N^{(2)}_{ \ell } = N^{(1)}_{ \ell } = N_{ \ell }$ for every $\ell \in \mylight( N )$. However, for each heavy class $\ell \in \myheavy( N )$, we will define $N^{(2)}_{ \ell }$ by eliminating a certain number of items from $N^{(1)}_{ \ell }$. To formalize this notion, let $\Delta_{ \ell } = N^{(1)}_{ \ell } - \frac{ 1 }{ \eps }$ be the number of items picked by $N^{(1)}$ from the profit class ${\cal P}_{ \ell }$ beyond the first $1 / \eps$ items. Then, $N^{(2)}_{ \ell }$ is obtained from $N^{(1)}_{ \ell }$ by eliminating its last $\lceil 2\eps \cdot \Delta_{ \ell } \rceil$ items, that is, $N^{(2)}_{ \ell } = N^{(1)}_{ \ell } - \lceil 2\eps \cdot \Delta_{ \ell } \rceil$.

\paragraph{Is truncation class-preserving?} Before turning our attention to specific structural properties of the newly obtained vector $N^{(2)}$, it is worth pointing out that, similarly to up-rounding, the truncation operation is class-preserving. To verify this claim, we first observe that for $\ell \in \mylight(N^{(1)})$, one clearly has $N^{(2)}_{ \ell } = N^{(1)}_{ \ell }$ by definition, implying that $\ell \in \mylight(N^{(2)})$. In addition, to argue that each class $\ell \in \myheavy(N^{(1)})$ remains heavy for $N^{(2)}$, we should prove that none of the first $1 / \eps$ items in $N^{(1)}_{ \ell }$ will be eliminated. Put differently, the number of eliminated items, $\lceil 2\eps \cdot \Delta_{ \ell } \rceil$, does not exceed $N^{(1)}_{ \ell } - \frac{ 1 }{ \eps } = \Delta_{ \ell }$. For this purpose, we know that $\Delta_{\ell} \geq 1$, as class $\ell$ is heavy. Now, when $\Delta_{\ell} = 1$, we have $\lceil 2\eps \cdot \Delta_{ \ell } \rceil = \lceil 2\eps \rceil = 1$ since $\eps < 1/4$, and the claim holds. When $\Delta_{\ell} \geq 2$, we have $\lceil 2\eps \cdot \Delta_{ \ell } \rceil \leq \frac{ \Delta_{ \ell } }{ 2 } + 1 \leq \Delta_{ \ell }$ for the same reason. To summarize,  we have $\mylight( N^{(2)} ) = \mylight( N^{(1)} ) = \mylight( N )$ and $\myheavy( N^{(2)} ) = \myheavy( N^{(1)} ) = \myheavy( N )$.

\paragraph{Key properties.} The next lemma, whose proof is provided in Appendix~\ref{subsec:proof_lem_properties_truncation}, summarizes the main properties of our construction.

\begin{lemma} \label{lem:properties_truncation}
The truncation operation satisfies the following properties:
\begin{enumerate}[label=(\alph*)]
\item Monotonicity: $N^{(2)} \leq N^{+(2)}$ for every $N \leq N^+$. \label{lem:truncation_monotone}

\item Weight restoration: $w( N^{(2)} ) \leq w( N )$. \label{lem:truncation_weight}

\item Items picked out of heavy classes: $N^{(2)}_{ \ell } \geq (1 - 2\eps ) \cdot N_{ \ell }$ for every $\ell \in \myheavy( N )$. \label{lem:truncation_picks_heavy}
\end{enumerate}
\end{lemma}

\subsection{Analysis: State space size and performance guarantee} \label{subsec:analysis}

\paragraph{Main analytical questions.} Following the discussion in Sections~\ref{subsec:overview}-\ref{subsec:truncation}, the family of utilization vectors $\familysmall = \{ N^{(2)} : N \in \familyall \}$ is now concrete and well-defined. Nevertheless, a number of basic questions still have to be addressed:
\begin{enumerate}
\item What is the cardinality of $\familysmall$?

\item What is the overall running time of the dynamic program $\tilde{F}_{\familysmall}$?

\item What is the optimality loss of $\opt_{\familysmall}$ relative to $\opt$?
\end{enumerate}
The remainder of this section provides a detailed treatment for each of these questions.

\paragraph{Answer 1: Bounding $|{\familysmall}|$.} Even though we have shown in Section~\ref{subsec:exact_dp} that the original collection $\familyall$ of utilization vectors may generally be of size $\Omega( n^{ \Omega(|{\cal L}|) } ) = \Omega( n^{ \Omega( \frac{ 1 }{ \eps } \log \frac{ n \rho }{ \eps } ) } ) $, it turns out that $\familysmall$ is considerably smaller, as stated in the next claim. 

\begin{lemma} \label{lem:bound_usmall_size}
$| \familysmall | = O ( (n \rho)^{ \tilde{O}( 1 / \eps^2 ) } \cdot \log ( \frac{ w_{\max} }{ w_{\min} } ) )$.
\end{lemma}
\proof{\em Proof.}
To establish the desired claim, the crucial observation is that, for any $N \in \familyall$, its truncated up-rounded image $N^{(2)} \in \familysmall$ is uniquely determined, given the following pieces of information:
\begin{enumerate}
\item {\em The identity of $N^{(2)}$'s light and heavy classes.} There are only $2^{ |{\cal L}| }$ possible ways of assigning a light/heavy label to each of the classes $\{ {\cal P}_{\ell} \}_{ \ell \in {\cal L} }$.

\item {\em The number of items picked by $N^{(2)}$ out of each light class.} In regard to these features, since $N^{(2)}_{ \ell } \leq \frac{ 1 }{ \eps }$ for all $\ell \in \mylight( N^{(2)} )$, there are only $O( ( \frac{ 1 }{ \eps } )^{ |\mylight( N^{(2)} )| } ) = O( ( \frac{ 1 }{ \eps } )^{ O( |{\cal L}|) } )$ possible values that can jointly be taken by $(N^{(2)}_{ \ell })_{\ell \in \mylight( N^{(2)} )}$.

\item {\em The number of items picked by $N^{(2)}$ out of each heavy class.} Deriving an upper bound on the number of possible values jointly taken by $(N^{(2)}_{ \ell })_{\ell \in \myheavy( N^{(2)} )}$ is somewhat more involved. For readability purposes, we separately show in Lemma~\ref{lem:bound_values_heavy} that this quantity is only $O( 2^{ O( |{\cal L}| / \eps ) } \cdot  \log ( \frac{ n w_{\max} }{ w_{\min} }) )$.
\end{enumerate}
By combining the above remarks, we have just shown that
\[ \left| \familysmall \right| ~~=~~ O \left( 2^{ |{\cal L}| } \cdot \left( \frac{ 1 }{ \eps } \right)^{ O( |{\cal L}|) } \cdot 2^{ O( |{\cal L}| / \eps ) } \cdot  \log \left( \frac{ n w_{\max} }{ w_{\min} } \right) \right) ~~=~~ O \left( (n \rho)^{ \tilde{O}( 1 / \eps^2 ) } \cdot \log \left( \frac{ w_{\max} }{ w_{\min} } \right) \right) \ , \]
where the second equality follows from plugging in $| {\cal L} | = O( \frac{ 1 }{ \eps } \log \frac{ n \rho }{ \eps } )$. \halmos
\endproof

\begin{lemma} \label{lem:bound_values_heavy}
The number of values jointly taken by $(N^{(2)}_{ \ell })_{\ell \in \myheavy( N^{(2)} )}$ is $O( 2^{ O( |{\cal L}| / \eps ) } \cdot  \log ( \frac{ n w_{\max} }{ w_{\min} }) )$.
\end{lemma}
\proof{\em Proof.}
A close inspection of how truncations are defined in Section~\ref{subsec:truncation} reveals that $N^{(2)}_{ \ell }$ is uniquely determined by $N^{(1)}_{ \ell }$. Therefore, since $\myheavy( N^{(1)} ) = \myheavy( N^{(2)} )$, we can instead focus on the possible values for $(N^{(1)}_{ \ell })_{\ell \in \myheavy( N^{(1)} )}$. In turn, the way up-roundings are defined in Section~\ref{subsec:up_rounding} guarantees these values to be uniquely determined by $\mypower_2 [ \frac{ \eps }{ |{\cal L}| } \cdot W_{ \myheavy }( N ) ]$ and by $(\mu_{ \ell } )_{ \ell \in \myheavy( N ) }$.

With respect to the first parameter, the quantity $W_{ \myheavy }( N ) = \sum_{ \ell \in \myheavy( N ) } w ( {\cal P}_{ \ell } [ \frac{ 1 }{ \eps } + 1, N_{\ell} ] )$ clearly resides within the interval $[w_{\min}, n  w_{\max}]$, where $w_{\min} = \min_{i \in [n]} w_i$ and $w_{\max} = \max_{i \in [n]} w_i$ stand for the minimum and maximum weights of any item, respectively. Consequently, $\mypower_2 [ \frac{ \eps }{ |{\cal L}| } \cdot W_{ \myheavy }( N ) ]$ is an integer power of 2 contained in the interval $[ \frac{ \eps }{ |{\cal L}| } \cdot w_{\min}, \frac{ 2\eps }{ |{\cal L}| } \cdot n w_{\max}]$, and there are only $O( \log (\frac{ n w_{\max} }{ w_{\min} }) )$ such powers. Now, as far as the values  $(\mu_{ \ell } )_{ \ell \in \myheavy( N ) }$ are concerned, we exploit the next claim, which places an upper bound on their summation. To avoid deviating from the broader argument, we provide the proof in Appendix~\ref{app:proof_clm_ub_sum_mu}.

\begin{claim} \label{clm:ub_sum_mu}
$\sum_{ \ell \in \myheavy( N ) } \mu_{ \ell } \leq \frac{ 3 }{ 2 } \cdot \frac{ |{\cal L}| }{ \eps }$.
\end{claim}

Now, basic counting ideas show that the number of different solutions in non-negative integers to the inequality $\sum_{ \ell \in \myheavy( N ) } \mu_{ \ell } \leq \frac{ 3 }{ 2 } \cdot \frac{ |{\cal L}| }{ \eps }$ is precisely
\[ \sum_{k = 0}^{ \lfloor \frac{ 3 }{ 2 } \cdot \frac{ |{\cal L}| }{ \eps } \rfloor } \binom{ | \myheavy( N ) | + k - 1 }{ k } ~~\leq~~ \sum_{k = 0}^{ \lfloor \frac{ 3 }{ 2 } \cdot \frac{ |{\cal L}| }{ \eps } \rfloor } \binom{ | {\cal L} | + \lfloor \frac{ 3 }{ 2 } \cdot \frac{ |{\cal L}| }{ \eps } \rfloor }{ k } ~~\leq~~ 2^{ | {\cal L} | + \lfloor \frac{ 3 }{ 2 } \cdot \frac{ |{\cal L}| }{ \eps } \rfloor } ~~=~~ 2^{ O( |{\cal L}| / \eps ) } \ . \halmos \]
\endproof

\paragraph{Answer~2: Running time.} We now turn our attention to evaluate the running time incurred by the approximate dynamic program $\tilde{F}_{\familysmall}$. For this purpose,
the first observation is that the reduced family of utilization vectors $\familysmall$ can be constructed in $O(|\familysmall| \cdot |{\cal I}|^{ O(1) } )$ time. As argued within the proof of Lemma~\ref{lem:bound_usmall_size}, each truncated vector $N^{(2)} \in \familysmall$ is uniquely determined by: (1)~The identity of light and heavy classes; (2)~The precise number of items picked by $N^{(2)}$ out of each light class; and (3)~The number of items picked by $N^{(2)}$ out of each heavy class, as governed by the parameters $W_{\myheavy}(N)$ and $(\mu_{ \ell } )_{ \ell \in \myheavy( N ) }$. Following our proof, it is not difficult to verify that all configurations of these parameters can be enumerated over in $O(|\familysmall| \cdot |{\cal I}|^{ O(1) } )$ time.

Now, for the purpose of solving the recursive equations~\eqref{eq:ADP} through which $\tilde{F}_{\familysmall}$ is defined, at each state $(t,N_t)$ of this dynamic program, the optimal action to be taken can straightforwardly be identified in $O(|{\cal L}|\cdot|\familysmall|)$ time. All in all, recalling that $| \familysmall | = O ( (n \rho)^{ \tilde{O}( 1 / \eps^2 ) } \cdot \log ( \frac{ w_{\max} }{ w_{\min} } ) )$ by Lemma~\ref{lem:bound_usmall_size}, the running time stated in Theorem~\ref{thm:main_bounded_lambda} is obtained by noting that the overall number of elementary operations performed is
\[ O \left( |\familysmall| \cdot |{\cal I}|^{ O(1) } + T\cdot |\familysmall|^2 \cdot |{\cal L}| \right) = O \left( (n \rho)^{ \tilde{O}( 1 / \eps^2 ) } \cdot |{\cal I}|^{ O(1) } \right) \ . \]

\paragraph{Answer 3: Performance guarantee.} We proceed by analyzing the approximate dynamic program $\tilde{F}_{\familysmall}$ in terms of its profit-wise performance. In the next claim, we bound the optimality loss incurred by replacing our original state space with the one induced by utilization vectors within $\familysmall$, thereby completing the proof of Theorem~\ref{thm:main_bounded_lambda}.

\begin{lemma} \label{lem:opt-familysmall}
$\opt_{\familysmall} \geq (1-2\eps)\cdot\opt$.
\end{lemma}
\proof{\em Proof.}
Recall that $\opt$ is given by the maximal value of $F(T, N_T) $ attained by a utilization vector $N_T \in \familyall$ for which $w(N_T) \leq w( S^*_T )$. We denote the sequence of states traversed by the original dynamic program $F$ to attain this value by $(0,N_0^*), (1, N_1^*), \ldots, (T, N_T^*)$, noting that $N_0^*= 0$.

The crucial claim we exploit is that, when each of the above-mentioned utilization vectors is replaced by its truncated up-rounding, the sequence of states $(0,0), (1, N_1^{*(2)}), \ldots, (T, N_T^{*(2)})$ we obtain is necessarily feasible along the approximate program $\tilde{F}_{\familysmall}$. To understand why this claim holds, we have by definition $N_t^{*(2)} \in \familysmall$ for every $t \in [T]_0$, and therefore, it remains to explain why $N_{t-1}^{*(2)}$ is a feasible solution to the structural constraints of the recursive equations~\eqref{eq:ADP}, through which $\tilde{F}_{\familysmall}(t, N_{t}^{*(2)})$ is defined. Specifically, these constraints stipulate that $N_{t-1}^{*(2)} \leq N_t^{*(2)}$ and $w(N_T^{*(2)}) \leq W_t$. For this purpose, note that $w( N_t^{*(2)} ) \leq w( N_t^* ) \leq W_t$, where the first inequality follows from Lemma~\ref{lem:properties_truncation}\ref{lem:truncation_weight} and the second inequality is a constraint of the dynamic program $F$. Along the same lines, since $N_{t-1}^* \leq N_t^*$ is a concurrent constraint of $F$, we have $N_{t-1}^{*(2)} \leq N_t^{*(2)}$ as well, by Lemma~\ref{lem:properties_truncation}\ref{lem:truncation_monotone}.

Now, recalling that $\opt_{\familysmall}$ is the maximal value of $\tilde{F}_{\familysmall}(T,N_T)$ attained by a utilization vector $N_T \in \familysmall$ for which $w(N_T) \leq w( S^*_T )$, we have just shown that $\opt_{\familysmall} \geq \tilde{F}_{\familysmall}(T, N_T^{*(2)})$, since $w(N_T^{*(2)}) \leq w(N_T^{*})\leq w( S^*_T )$, where the first inequality follows from Lemma~\ref{lem:properties_truncation}\ref{lem:truncation_weight}.  Thus, to conclude the proof, it suffices to argue that $\tilde{F}_{\familysmall}(T, N_T^{*(2)})\geq (1 - 2\eps) \cdot \opt$.  For this purpose, note that
\begin{eqnarray*}
\tilde{F}_{\familysmall}(T, N_T^{*(2)}) & = & \sum_{t \in [T]} \left( \tilde{F}_{\familysmall}(T, N_t^{*(2)}) - \tilde{F}_{\familysmall}(T, N_{t-1}^{*(2)})\right) \\
& \geq & \sum_{t \in [T]} \left( \lambda^{ t+ } \cdot \sum_{\ell \in {\cal L}} (1 + \eps)^{ \ell } \cdot \left( N_{t,\ell}^{*(2)} - N_{t-1,\ell}^{*(2)} \right) \right) \\
& = & \sum_{\ell \in {\cal L}} \Bigg( (1+\eps)^{ \ell } \cdot \sum_{t \in [T]} \lambda^{ t+ } \left( N_{t,\ell}^{*(2)} - N_{t-1,\ell}^{*(2)} \right) \Bigg) \\
& = & \sum_{\ell \in {\cal L}} \Bigg( (1+\eps)^{ \ell } \cdot \sum_{t \in [T]} \lambda_t \cdot N_{t,\ell}^{*(2)} \Bigg) \ ,
\end{eqnarray*}
where the inequality above holds since $N_{t-1}^{*(2)}$ is one of the feasible actions within the dynamic programming equation~\eqref{eq:ADP} through which $\tilde{F}_{\familysmall}(T, N_t^{*(2)})$ is defined. On the other hand, using nearly identical transitions, we have
\begin{eqnarray*}
\opt & = & F\left(T,N^*_T\right)\\
& = &
\sum_{t \in [T]} \left(  F\left(t,N^*_t\right) -  F\left(t-1,N^*_{t-1}\right) \right) \\
& = & \sum_{t \in [T]} \left( \lambda^{ t+ } \cdot \sum_{\ell \in {\cal L}} (1 + \eps)^{ \ell } \cdot \left( N_{t,\ell}^* - N_{t-1,\ell}^* \right) \right) \\
& = & \sum_{\ell \in {\cal L}} \Bigg( (1+\eps)^{ \ell } \cdot \sum_{t \in [T]} \lambda_t \cdot N_{t,\ell}^* \Bigg) \ .
\end{eqnarray*}
Consequently, we show that $\tilde{F}_{\familysmall}(T, N_T^{*(2)})\geq (1 - 2\eps) \cdot \opt$ by comparing between these quantities on a term-by-term basis. Specifically, as mentioned in Section~\ref{subsec:truncation}, we have $N_{t,\ell}^{*(2)} = N_{t,\ell}^{*}$ for every $\ell \in \mylight( N_{t}^{*} )$. In addition, by Lemma~\ref{lem:properties_truncation}\ref{lem:truncation_picks_heavy}, we have $N_{t,\ell}^{*(2)} \geq (1 - 2\eps ) \cdot N_{t,\ell}^{*}$ for every $\ell \in \myheavy( N_{t}^{*} )$. \halmos
\endproof 
\section{PTAS for General Instances} \label{sec:alg_general}

In this section, we devise a polynomial-time approximation scheme for arbitrarily structured instances of the incremental knapsack problem by exploiting our algorithm for the $\lambda^+$-bounded setting as a subroutine. The main result we establish is summarized in Theorem~\ref{thm:general}, which is restated here for readability purposes.

\mainresult*

\subsection{Technical overview} \label{subsec:overview_general}

To establish our main result, we present an approximate reduction to the $\lambda^+$-bounded setting, previously examined in Section~\ref{sec:alg_bounded}. For this purpose, we develop a dynamic programming approach that decomposes arbitrarily structured instances of the incremental knapsack problem into mutually disjoint sub-problems, each focusing on a restricted subset of time periods, collectively referred to as a cluster. These sub-problems are instances of the inverse incremental knapsack problem, shown to satisfy the $\lambda^+$-bounded property with an $O((n/\eps)^{O(1/\eps)})$ ratio between their extremal $\lambda^+$ values. Consequently, we leverage our algorithm for the $\lambda^+$-bounded setting as a subroutine to efficiently approximate each sub-problem. In what follows, we outline the main technical ingredients of our approach.

\paragraph{Clustering time periods (Section~\ref{subsec:clustering}).} Given a general instance of the incremental knapsack problem, we start-off  by defining clusters of time periods. Each such cluster is formed by an interval of successive time periods, such that within clusters, $\lambda^+$ values may differ by a factor of $O((n/\eps)^{O(1/\eps)})$. Moreover, these clusters are ``well-separated'', in the sense that there is a large gap between their underlying $\lambda^+$ values, a notion that will be made formal later on.

\paragraph{Near-optimal uncrossing-stars solutions (Section~\ref{subsec:uncrossing}).} Building on this construction, we inspect the relationship between clusters and profit classes within near-optimal solutions. Specifically, we examine the bipartite graph induced by such solutions, having cluster vertices on one side and profit class vertices on the other, whose edges represent whether or not at least one item in a given profit class is introduced within a given cluster of time periods. Our main structural result proves the existence of a near-optimal solution that induces a highly structured bipartite graph, taking the form of disjoint uncrossing-stars. Informally, each connected component of this graph is a star centered at a time period cluster, which is joined by edges to profit classes that follow a very specific ordering property.

\paragraph{Dynamic programming decomposition (Sections~\ref{subsec:dp-stars}-\ref{subsec:analysis-stars}).} Finally, we leverage the above-mentioned existence result to obtain a polynomial-time approximation scheme by showing that uncrossing-stars solutions can be efficiently computed. Specifically, we argue that one can identify a near-optimal uncrossing-stars solution via a dynamic programming based decomposition. Interestingly, the recursive equations involved in detecting each star-shaped solution boil down to handling a $\lambda^+$-bounded sub-problem, allowing us to utilize the approximation scheme of Section~\ref{sec:alg_bounded} as a subroutine.

\subsection{Clustering time periods} \label{subsec:clustering}

\paragraph{Defining clusters.} We define the sequence of time period intervals $I_1 = \{ t \in [T] : \lambda^{t+} \in ( \frac{ \eps }{ n } \cdot \lambda^{1+}, \lambda^{1+}] \}$, $I_2 = \{ t \in [T] : \lambda^{t+} \in ( ( \frac{ \eps }{ n } )^2 \cdot \lambda^{1+}, \frac{ \eps }{ n } \cdot \lambda^{1+}] \}$, so on and so forth. Assuming without loss of generality that $1 / \eps$ is an integer, we randomly pick $\xi \sim U \{ 0, \ldots, \frac{ 1 }{ \eps } - 1 \}$. With respect to this choice, any interval $I_m$ with $(m \mod \frac{ 1 }{ \eps }) = \xi$ is said to be bad, while all other intervals are good. We use ${\cal C}_1$ to denote the set of time periods belonging to the (good) intervals appearing prior to the first bad interval (i.e., $I_1, \ldots, I_{\xi-1}$). Then, ${\cal C}_2$ will denote the set of time periods in the (good) intervals appearing between the first and second bad intervals (i.e., $I_{\xi+1}, \ldots, I_{\xi + \frac{ 1 }{ \eps } - 1}$), so on and so forth, up until the last non-empty such set, ${\cal C}_M$. We refer to ${\cal C}_1, \ldots, {\cal C}_M$ as time period clusters, noting that within a single cluster, $\lambda^{t+}$ values differ by a factor of at most $(n/\eps)^{(1/\eps)-1}$ and that the overall number of clusters is $M = O( \log_{ (n/\eps)^{1/\eps} } ( \frac{ \lambda^{1+} }{ \lambda^{T+} } ) ) = O( | {\cal I} |^{ O(1) } )$.

\paragraph{The cost of ignoring bad intervals.} Let ${\cal S}^* = \langle S_1^*, \ldots, S_T^* \rangle$ be a fixed optimal solution, with a profit of $\objfunc( {\cal S}^* ) = \sum_{i \in S_T^*} \objfunc_i ( {\cal S}^* )$, written according to representation~\eqref{eqn:profit_decomp}. Suppose we create a new feasible solution ${\cal S}_{ \xi }$ by modifying ${\cal S}^*$ as follows: In each period $t$ belonging to a bad interval, we do not introduce new items. In other words, all items in $S_t^* \setminus S_{t-1}^*$ are deleted from $S_t^*, \ldots, S_T^*$. Equivalently, ${\cal S}_{ \xi }$ is constructed by leaving only items that were introduced in good time periods. Clearly, due to the randomness in choosing $\xi$, this is a random solution. The next lemma shows that, in expectation over the choice of $\xi$, the profit of ${\cal S}_{ \xi }$ is near optimal. As a result, at least one choice deterministically guarantees the bound stated below.

\begin{lemma} \label{lem:expected_profit_cluster}
$\expartwo{\xi}{ \objfunc( {\cal S}_{ \xi } ) } = (1 - \eps) \cdot \objfunc( {\cal S}^* )$.
\end{lemma}
\proof{\em Proof.}
To prove the desired claim, we first observe that each item $i \in S_T^*$ contributes $\objfunc_i ( {\cal S}^* )$ toward $\objfunc( {\cal S}_{ \xi } )$ when the time period $\nu(i,{\cal S}^*)$ in which it is introduced resides in a good interval, and does not contribute at all otherwise. Therefore,
\begin{eqnarray*}
\extwo{\xi}{ \objfunc( {\cal S}_{ \xi } ) } & = & \extwo{\xi}{ \sum_{i \in S_T^*} \objfunc_i \left( {\cal S}^* \right) \cdot \bbI \left[ \nu(i,{\cal S}^*) \text{ in good interval} \right] } \\
& = & \sum_{i \in S_T^*} \objfunc_i \left( {\cal S}^* \right) \cdot \prtwo{ \xi }{ \nu(i,{\cal S}^*) \text{ in good interval} } \\
& = & (1 - \eps) \cdot \sum_{i \in S_T^*} \objfunc_i \left( {\cal S}^* \right) \\
& = & (1 - \eps) \cdot \objfunc \left( {\cal S}^* \right) \ ,
\end{eqnarray*}
where the third equality holds since any interval is good with probability $1-\eps$. \halmos
\endproof

\subsection{Uncrossing-stars solutions}  \label{subsec:uncrossing}

By enumerating over the collection of values $0, \ldots, \frac{ 1 }{ \eps } - 1$, we may assume without loss of generality that a particular choice of $\xi$ for which $\objfunc( {\cal S}_{ \xi } ) \geq (1 - \eps) \cdot \objfunc( {\cal S}^* )$ is known in advance. Now let $\hat{\cal S} = \langle \hat{S}_1, \ldots, \hat{S}_T \rangle$ be a maximum-profit solution that only introduces items in time periods belonging to good intervals, where ``good'' and ``bad'' are defined with respect to $\xi$. Clearly, $\objfunc( \hat{\cal S} ) \geq \objfunc( {\cal S}_{ \xi } ) \geq (1 - \eps) \cdot \objfunc( {\cal S}^* )$.

\paragraph{The solution graph.} From this point on, we focus our attention on solutions ${\cal S} = \langle S_1, \ldots, S_T \rangle$ that only make use of time periods within good intervals. For each such solution, we define an undirected bipartite graph $G( {\cal S} )$ as follows:
\begin{itemize}
\item Vertices: On one side, we have the time period clusters ${\cal C}_1, \ldots, {\cal C}_M$. On the other side, we have the profit classes ${\cal P}_0, \ldots, {\cal P}_L$, where $L$ stands for the maximal index of a non-empty profit class.

\item Edges: There is an edge $( {\cal C}_m, {\cal P}_{\ell} )$ if and only if at least one item belonging to the profit class ${\cal P}_{\ell}$ is introduced by ${\cal S}$ in a time period belonging to ${\cal C}_m$.
\end{itemize}

\paragraph{Post-filtering graph properties.} We proceed by filtering items whose profit contribution is negligible. Specifically, let $\tilde{\cal S} = \langle \tilde{S}_1, \ldots, \tilde{S}_T \rangle$ be the solution obtained from $\hat{\cal S} = \langle \hat{S}_1, \ldots, \hat{S}_T \rangle$ by eliminating all items $i \in \hat{S}_T$ for which $\objfunc_i ( \hat{\cal S} ) \leq \frac{ \eps }{ n } \cdot \objfunc ( \hat{\cal S} )$; it is easy to verify that $\objfunc ( \tilde{\cal S} ) \geq (1 - \eps) \cdot \objfunc ( \hat{\cal S} )$. The crucial claim is that this filtering procedure results in a greatly simplified graph $G( \tilde{\cal S} )$. Lemma~\ref{lem:solution_structure} below shows that this graph is a collection of uncrossing stars, formally defined as follows:
\begin{enumerate}
\item {\em ${\cal P}$-degrees}: Each of the vertices ${\cal P}_0, \ldots, {\cal P}_L$ has a degree of at most $1$.

\item {\em Uncrossing}: There are no crossing edges, namely, a pair of edges $( {\cal C}_{m_1}, {\cal P}_{\ell_1} )$ and $( {\cal C}_{m_2}, {\cal P}_{\ell_2} )$ with $\ell_1 < \ell_2$ and $m_1 > m_2$.
\end{enumerate}
Put differently, property~1 implies that every connected component of $G( \tilde{\cal S} )$ is star-shaped, centered at a time period cluster. Property~2 imposes that, for any two such stars, say centered at ${\cal C}_{m_1}$ and ${\cal C}_{m_2}$ with $m_1 < m_2$, all profit classes within ${\cal C}_{m_1}$'s star have lower indices than those within ${\cal C}_{m_2}$'s star, which will be instrumental for algorithmic purposes, as shown in Sections~\ref{subsec:dp-stars}-\ref{subsec:analysis-stars}.

\begin{lemma} \label{lem:solution_structure}
$G( \tilde{\cal S} )$ is a collection of uncrossing stars.
\end{lemma}
\proof{\em Proof.}
In what follows, we prove the ${\cal P}$-degrees property, noting that the uncrossing property can be established via nearly identical arguments and is omitted to avoid duplicity. To arrive at a contradiction, suppose that $G( \tilde{\cal S} )$ has some vertex ${\cal P}_{\ell}$ with degree at least $2$, and let $( {\cal C}_{m_1}, {\cal P}_{\ell} )$ and $( {\cal C}_{m_2}, {\cal P}_{\ell} )$ be two of its adjacent edges; without loss of generality, $m_1 < m_2$. By construction of $G( \tilde{\cal S} )$, at least one item $i_1 \in {\cal P}_{\ell}$ is introduced by $\tilde{\cal S}$ in a time period belonging to ${\cal C}_{m_1}$, implying that the pre-filtered solution $\hat{\cal S}$ is associated with a profit of
\begin{equation} \label{eqn:lb_obj_item1}
\objfunc ( \hat{\cal S} ) ~~\geq~~ \objfunc_{i_1} ( \hat{\cal S} ) ~~=~~ \objfunc_{i_1} ( \tilde{\cal S} ) ~~=~~ p_{i_1} \cdot \lambda^{ \nu(i_1,\tilde{\cal S}) + } ~~\geq~~ (1+\eps)^{ \ell } \cdot \lambda^{ \max {\cal C}_{m_1} + } \ ,
\end{equation}
where the last inequality holds since items in ${\cal P}_{\ell}$ have a uniform profit of $(1+\eps)^{\ell}$ and since $\nu(i_1,\tilde{\cal S}) \in {\cal C}_{m_1}$, over which the smallest $\lambda^{ +}$ value
is attained at $\lambda^{ \max {\cal C}_{m_1} + }$, as $\lambda^{1+} \geq \cdots \geq \lambda^{T+}$. Similarly, at least one item $i_2 \in {\cal P}_{\ell}$ is introduced in a time period belonging to ${\cal C}_{m_2}$, meaning that its contribution to the pre-filtered profit satisfies
\[ \objfunc_{i_2} ( \hat{\cal S} ) ~~=~~ \objfunc_{i_2} ( \tilde{\cal S} ) ~~=~~ p_{i_2} \cdot \lambda^{ \nu(i_2,\tilde{\cal S}) + } ~~\leq~~ (1+\eps)^{ \ell } \cdot \lambda^{ \min {\cal C}_{m_2} + } ~~\leq~~ (1+\eps)^{ \ell } \cdot \frac{ \eps }{ n } \cdot \lambda^{ \max {\cal C}_{m_1} + } ~~\leq~~ \frac{ \eps }{ n } \cdot \objfunc ( \hat{\cal S} ) \ , \]
where the second inequality holds since $\lambda^{ \min {\cal C}_{m_2} + } \leq
( \frac{ \eps }{ n } )^{ m_2 - m_1 } \cdot \lambda^{ \max {\cal C}_{m_1} + } \leq
\frac{ \eps }{ n }  \cdot \lambda^{ \max {\cal C}_{m_1} + }$, by construction of the time period clusters, and the third inequality is obtained by plugging-in the bound in~\eqref{eqn:lb_obj_item1}. We have just arrived at a contradiction: Since $\objfunc_{i_2} ( \hat{\cal S} ) \leq \frac{ \eps }{ n } \cdot \objfunc ( \hat{\cal S} )$, the item $i_2$ should have been eliminated as part of our filtering procedure, while constructing the solution $\tilde{\cal S}$. \halmos
\endproof

\subsection{The exact dynamic program}  \label{subsec:dp-stars}

In the remainder of this section, we develop a dynamic programming approach for computing a near-optimal uncrossing-stars solution by leveraging our algorithm for the $\lambda^+$-bounded setting.

\paragraph{State space.} Each state of our dynamic program is described by three parameters:
\begin{itemize}
\item The index $m$ of the current cluster, taking one of the values $0,1\ldots,M$.

\item The index $\ell_m$ of the current profit class, whose possible values are $-1,0,\ldots, L$.

\item The total profit $\varphi_m$ accumulated thus far. For the time being, $\varphi_m$ will be treated as a continuous parameter, restricted to the interval $[0,\lambda^{1+}\cdot \sum_{i\in [n]}p_i]$.
\end{itemize}
We mention in passing that $m = 0$ does not correspond to a concrete cluster, and we therefore define ${\cal C}_0 = \emptyset$ by convention. Similarly, $\ell_m = -1$ is an auxiliary value, and we set ${\cal P}_{-1} = \emptyset$ for completeness. These edge cases will be useful in prescribing the terminal conditions of our dynamic program. In addition, due to making use of a continuous profit parameter $\varphi_m$, the initial dynamic programming formulation we propose should not be viewed as a discrete algorithm, but rather as a characterization of optimal uncrossing-stars solutions. We will explain in Section~\ref{subsec:adp-stars} how to efficiently approximate this dynamic program through an appropriate discretization of $\varphi_m$.

\paragraph{Value function.} Letting $\tau_m = \max {\cal C}_m$, the value function $F(m, \ell_m, \varphi_m)$ stands for the minimum-possible total weight $w(S_{\tau_m})$ incurred by a $\tau_m$-period uncrossing-stars solution ${\cal S} = \langle S_1, \ldots, S_{\tau_m}\rangle$ that satisfies the following conditions:
\begin{enumerate}
\item {\em $[1,m]$-clustered periods:} Items may be introduced only within time periods belonging to one of the clusters ${\cal C}_1, \ldots, {\cal C}_m$. In other words, $S_t \setminus S_{t-1} = \emptyset$ for every $t \notin \biguplus_{\mu \in [m]} {\cal C}_{\mu}$.

\item {\em $[0,\ell_m]$-profit classes:} We are allowed to introduce only items that reside within the profit classes ${\cal P}_0, \ldots, {\cal P}_{\ell_m}$, meaning that $S_{\tau_m} \subseteq \biguplus_{\ell \in [\ell_m]_0} {\cal P}_{\ell}$.

\item {\em Total profit:} We are restricted to attain a profit of at least $\varphi_m$, or put differently, $\Psi({\cal S}) \geq \varphi_m$.
\end{enumerate}
A close inspection of this definition reveals that we have actually formulated a multi-cluster instance of the inverse incremental knapsack problem, whose feasibility set is further limited to uncrossing-stars solutions. We designate the latter instance by ${\cal I}_{ \mymulti }^{\varphi_m} [ m,\ell_m ]$, noting that $F(m, \ell_m, \varphi_m)$ corresponds to its optimum value. When ${\cal I}_{ \mymulti }^{\varphi_m} [ m,\ell_m ]$ is infeasible, we define $F(m, \ell_m, \varphi_m) = \infty$. Clearly, the optimal profit attained by an uncrossing-stars solution is precisely the maximal value $\varphi_M$ for which $F(M, L, \varphi_M) <  \infty$. As an immediate conclusion of Section~\ref{subsec:uncrossing}, the latter parameter can be related to the optimal profit as follows.

\begin{corollary} \label{cor:f_vs_opt}
$F(M, L, \varphi_M) <  \infty$ for any $\varphi_M \leq (1 - 2\eps) \cdot \objfunc( {\cal S}^* )$.
\end{corollary}

\paragraph{Optimal substructure.} In order to derive recursive equations for this value function, for every $\phi \geq 0$ and $\omega \geq 0$, let us define a single-cluster instance ${\cal I}_{ \mysingle }^{\phi,\omega} [ m,\ell^-,\ell^+ ]$ of the inverse incremental knapsack problem as follows:
\begin{itemize}
\item The underlying collection of items is given by those residing in the profit classes ${\cal P}_{\ell^-}, \ldots, {\cal P}_{\ell^+}$.

\item The time periods are only those within the cluster ${\cal C}_m$. Each such period $t$ is given an adjusted capacity of $[W_t - \omega]^+$.

\item The objective is to compute a solution ${\cal S} = \langle S_{ \min {\cal C}_m }, \ldots, S_{ \max {\cal C}_m } \rangle$ that minimizes $w( S_{ \max {\cal C}_m } )$ subject to the profit requirement $\Psi({\cal S}) \geq \phi$. Here, the profit function $\Psi$ is specialized for the time periods in question, namely, $\objfunc( {\cal S} ) = \sum_{t \in {\cal C}_m} \lambda_t \cdot p( S_t )$.
\end{itemize}
Moving forward, we will use $\opt( {\cal I}_{ \mysingle }^{\phi,\omega} [ m,\ell^-,\ell^+ ] )$ to designate the optimum value of ${\cal I}^{\phi,\omega} [ m,\ell^-,\ell^+ ]$, with the convention that $\opt( {\cal I}_{ \mysingle }^{\phi,\omega} [ m,\ell^-,\ell^+ ] ) = \infty$ when this instance is infeasible.

Now, to understand where such instances fit in, suppose we are given an optimal $\tau_m$-period uncrossing-stars solution ${\cal S} = \langle S_1, \ldots, S_{\tau_m}\rangle$ for the multi-cluster instance ${\cal I}_{ \mymulti }^{\varphi_m} [ m,\ell_m ]$, satisfying in particular $w( S_{\tau_m} ) = F(m,\ell_m,\varphi_m)$. Our recursive approach proceeds by decomposing ${\cal S}$ into two separate solutions:
\begin{itemize}
\item The restriction of ${\cal S}$ to time periods in lower-indexed clusters, ${\cal S}^{<m} = \langle S_{1}, \ldots, S_{\tau_{m-1}}\rangle$.

\item The marginal sequence of items introduced within cluster ${\cal C}_m$, given by ${\cal S}^m = \langle S_{ \min {\cal C}_m } \setminus S_{\tau_{m-1}}, \ldots, S_{ \max {\cal C}_m } \setminus S_{\tau_{m-1}} \rangle$.
\end{itemize}
Focusing on the restriction ${\cal S}^{<m}$, let $\varphi_{m-1} = \Psi({\cal S}^{<m})$ be the profit obtained by this solution, and let $\ell_{m-1} \in [ \ell_m ]_0$ be the maximum index of a profit class out of which at least one item is introduced, i.e., $\ell_{m-1} = \max \{ \ell \in [ \ell_m ]_0 : S_{\tau_{m-1}} \cap {\cal P}_{\ell} \neq \emptyset \}$. The next claim sheds light on the optimal substructure pertaining to our decomposition of ${\cal S}$ into ${\cal S}^{<m}$ and ${\cal S}^m$.

\begin{lemma} \label{lem:opt_structure}
The solution ${\cal S}^{<m}$ is optimal for the multi-cluster instance ${\cal I}_{ \mymulti }^{\varphi_{m-1}} [ m-1,\ell_{m-1} ]$, and ${\cal S}^{m}$ is optimal for the single-cluster instance  ${\cal I}_{ \mysingle}^{\varphi_m - \varphi_{m-1}, w(S^{<m}_{\tau_{m-1}})}[m,\ell_{m-1}+1,\ell_{m}]$.
\end{lemma}
\proof{\em Proof.}
We first note that, by definition, ${\cal S}^{<m}$ is a feasible $\tau_{m-1}$-period uncrossing-stars solution to the multi-cluster instance ${\cal I}_{ \mymulti }^{\varphi_{m-1}} [ m-1,\ell_{m-1} ]$. In addition, ${\cal S}^{m}$ is feasible for the single-cluster instance ${\cal I}_{ \mysingle}^{\varphi_m - \varphi_{m-1}, w(S^{<m}_{\tau_{m-1}})}[m,\ell_{m-1}+1,\ell_{m}]$, again by definition. Now, to arrive at a contradiction, suppose that at least one of these solutions is sub-optimal for its corresponding instance. Then, there are feasible solutions $\tilde{\cal S}^{<m}$ and $\tilde{\cal S}^{m}$ to ${\cal I}_{ \mymulti }^{\varphi_{m-1}} [ m-1,\ell_{m-1} ]$ and ${\cal I}_{ \mysingle}^{\varphi_m - \varphi_{m-1}, w(S^{<m}_{\tau_{m-1}})}[m,\ell_{m-1}+1,\ell_{m}]$, respectively, for which
\[ w(\tilde{S}^{<m}_{\tau_{m-1}}) + w(\tilde{S}^{m}_{\tau_m}) < w(S^{<m}_{\tau_{m-1}}) + w(S^{m}_{\tau_m}) \ . \]
Therefore, to conclude the proof, it suffices to show that the $\tau_m$-period uncrossing-stars solution $\tilde{\cal S}$ obtained by gluing $\tilde{\cal S}^{<m}$ and $\tilde{\cal S}^{m}$ together is feasible for the multi-cluster instance ${\cal I}_{ \mymulti }^{\varphi_m} [ m,\ell_m ]$. This finding contradicts the optimality of ${\cal S}$, since the inequality above shows that the total weight $w(\tilde{S}_{\tau_m}) = w(\tilde{S}^{<m}_{\tau_{m-1}}) + w(\tilde{S}^{m}_{\tau_m})$ is strictly smaller than the total weight $w(S_{\tau_m}) = w(S^{<m}_{\tau_{m-1}}) + w(S^{m}_{\tau_m})$. To establish the feasibility of $\tilde{\cal S}$ for ${\cal I}_{ \mymulti }^{\varphi_m} [ m,\ell_m ]$, we verify each of the conditions involved:
\begin{enumerate}
\item {\em $[1,m]$-clustered periods:} By definition, any item introduced by $\tilde{\cal S}$ is either introduced by $\tilde{\cal S}^{<m}$, implying that this event occurs at a time period belonging to one of the clusters ${\cal C}_1, \ldots, {\cal C}_{m-1}$, or by $\tilde{\cal S}^{m}$, in which case, this event occurs at a time period in ${\cal C}_m$.

\item {\em $[0,\ell_m]$-profit classes:} Along the same lines, all items introduced by $\tilde{\cal S}^{<m}$ reside within the profit classes ${\cal P}_0, \ldots, {\cal P}_{\ell_{m-1}}$, whereas those introduced by $\tilde{\cal S}^{m}$ reside within ${\cal P}_{\ell_{m-1}+1}, \ldots, {\cal P}_{\ell_m}$.

\item {\em Total profit:} Here, the important observation is that, since $\tilde{\cal S}$ introduces precisely the same items as $\tilde{\cal S}^{<m}$ and $\tilde{\cal S}^{m}$ do, at precisely the same time periods, its profit can be bounded as follows:
    \[ \Psi(\tilde{\cal S}) ~~=~~ \Psi(\tilde{\cal S}^{<m}) + \Psi(\tilde{\cal S}^{m}) ~~\geq~~ \varphi_{m-1} + (\varphi_m - \varphi_{m-1}) ~~=~~ \varphi_m \ . \halmos \]
\end{enumerate}
\endproof

In light of the preceding discussion, we have just shown that
\begin{eqnarray}
F\left(m,\ell_m,\varphi_m\right) & = & w\left(S_{\tau_m}\right)  \nonumber\\
& = & w(S^{<m}_{\tau_{m-1}}) + w\left(S^m_{\tau_m}\right)  \nonumber\\
& = & F\left(m-1,\ell_{m-1},\varphi_{m-1}\right) + \opt \Big( {\cal I}_{ \mysingle }^{\varphi_{m}-\varphi_{m-1},w(S^{<m}_{\tau_{m-1}})} [ m,\ell_{m-1} + 1,\ell_m ] \Big) \nonumber\\
& = & F\left(m-1,\ell_{m-1},\varphi_{m-1}\right) + \opt \Big( {\cal I}_{ \mysingle }^{ \varphi_{m} - \varphi_{m-1}, F( m-1,\ell_{m-1},\varphi_{m-1} )} [ m,\ell_{m-1} + 1,\ell_m ] \Big) \label{eq:decompo-stars}  \ .	
\end{eqnarray}
Here, the first equality holds since ${\cal S}$ is an optimal solution to ${\cal I}_{ \mymulti }^{\varphi_m} [ m,\ell_m ]$. The second equality is obtained by observing that  $S_{\tau_m}$ is a disjoint union of $S^{<m}_{\tau_{m-1}}$ and $S^m_{\tau_m}$, by definition. The third and fourth equalities are direct implications of Lemma~\ref{lem:opt_structure}.

\paragraph{Recursive equations.} By taking advantage of this structural insight, we can now argue that the value function $F$ satisfies the following recursive equation: \begin{eqnarray*}
F\left(m,\ell_m, \varphi_m\right)  = &\min\limits_{\ell_{m-1},{\varphi}_{m-1}} & \left\{ F\left(m-1,\ell_{m-1},\varphi_{m-1}\right) + \opt \left( {\cal I}_{ \mysingle }^{ \phi, \omega} [ m,\ell_{m-1} + 1,\ell_m ] \right) \right\} \\
&{\mathrm{s.t.}} & \ell_{m-1} \in [\ell_m]_0 \ , \ \varphi_{m-1} \in [0, \varphi_m] \\
& &\phi = \varphi_{m}-\varphi_{m-1} \ , \\
& & \omega = F\left(m-1,\ell_{m-1},\varphi_{m-1}\right) \ .
\end{eqnarray*}
It is worth mentioning that $\phi$ and $\omega$ are not decision variables in this formulation; they simply serve as shorthand notation, to avoid having $\opt ( {\cal I}_{ \mysingle }^{ \varphi_{m}-\varphi_{m-1}, F( m-1, \ell_{m-1}, \varphi_{m-1})} [ \cdots ] )$ in the objective function. To better understand these recursive equations, note that decomposition~\eqref{eq:decompo-stars} proves the existence of at least one solution $(\ell_{m-1},{\varphi}_{m-1})$ to the maximization problem on the right hand side with an objective value of $F(m,\ell_m, \varphi_m)$. In parallel, for any choice of $\ell_{m-1} \in [\ell_m]_0$ and $\varphi_{m-1} \in [0, \varphi_m]$, their corresponding objective value provides an upper bound on $F(m,\ell_m, \varphi_m)$. Indeed, we can simply glue together the optimal solutions to ${\cal I}_{ \mymulti }^{\varphi_{m-1}} [ m-1,\ell_{m-1} ]$ and ${\cal I}_{ \mysingle}^{\phi,\omega}[m,\ell_{m-1}+1,\ell_{m}]$, thus forming a feasible solution to the multi-cluster instance ${\cal I}_{ \mymulti }^{\varphi_{m}} [ m,\ell_{m} ]$, implying that
\begin{eqnarray*}
& & F\left(m-1,\ell_{m-1},\varphi_{m-1}\right) + \opt \left( {\cal I}_{ \mysingle }^{ \phi, \omega} [ m,\ell_{m-1} + 1,\ell_m ] \right) \\
& & \quad =~ \opt \left({\cal I}_{ \mymulti }^{\varphi_{m-1}} \left[ m-1,\ell_{m-1} \right]\right) +  \opt \left({\cal I}_{ \mysingle}^{\phi,\omega}\left[m,\ell_{m-1}+1,\ell_{m}\right]\right)	\\
& & \quad \geq~ \opt \left({\cal I}_{ \mymulti }^{\varphi_{m}} \left[ m,\ell_{m} \right]\right) \\
& & \quad =~ F\left(m,\ell_m,\varphi_m\right) \ ,
\end{eqnarray*}
The terminal states of the dynamic program $F$ correspond to $m = 0$ or $\ell_{m} = -1$, in which case $F(m,\ell_m,0) = 0$ by definition and $F(m,\ell_m,\varphi_m) = \infty$ for $\varphi_m > 0$. Finally, from a computational standpoint, beyond the continuity of the profit parameters, our objective function involves an exact solution to single-cluster instance ${\cal I}_{ \mysingle }^{ \phi, \omega} [ m,\ell_{m-1} + 1,\ell_m ]$. To address this computational challenge, we  utilize the approximation scheme developed in Section~\ref{sec:alg_bounded}, as explained in the next section.

\subsection{The approximate program} \label{subsec:adp-stars}

With the above-mentioned obstacles in mind, we turn our attention to devising a polynomial-time approximation for the exact formulation $F$. To formalize this notion, we introduce two alterations of the original dynamic program:
\begin{itemize}
\item {\em Profit discretization}. In order to discretize the required profit $\varphi_m$, we restrict this parameter to taking values within the set
    \begin{equation} \label{eqn:def_phi}
    \Phi= \left\{0\right\} \cup \left\{\left(1+\frac{\eps}{M}\right)^k \cdot \delta: 0 \leq k \leq \left\lceil \frac{M}{\eps} \cdot \log\left(\frac{M}{\eps} \cdot \frac{\lambda^{1+}}{\lambda_T}\right)\right\rceil \right\} \ ,
    \end{equation}
    where $\delta = \frac{\eps}{M} \cdot\lambda_T\cdot p_{ \max }$. One can easily verify that this set is of cardinality $|\Phi| = O(\frac{1}{\eps}\log \frac{1}{\eps}\cdot |{\cal I}|^{O(1)} )$, by recalling that $M = O( |{\cal I}|^{O(1)} )$, as shown in Section~\ref{subsec:clustering}. From this point on, we use ${\cal D}$ to designate the discrete state space, in which $\varphi_m$ is restricted to $\Phi$, meaning that
    \[ {\cal D} = \left\{ ( m,\ell_m, \varphi_m ) : m \in \{ 0, 1, \ldots, M \}, \ell_m \in \{ -1, 0, \ldots, L \}, \varphi_m \in \Phi \right\} \ . \]

\item {\em Approximating single-cluster problems.} In addition, rather than shooting for an optimal solution to single-cluster instances of the form ${\cal I}_{ \mysingle }^{ \phi, \omega} [ m,\ell_{m-1} + 1,\ell_m ]$, we utilize the approximation scheme proposed in Theorem~\ref{thm:main_bounded_lambda} as a subroutine. Consequently, we are guaranteed to obtain in $O((n \rho)^{ \tilde{O}( 1 / \eps^2 ) } \cdot |{\cal I}|^{ O(1) } )$ time a super-optimal solution, i.e., one having a total weight of
    \begin{equation} \label{eqn:opt_vs_tildeopt}
    \widetilde{\opt} \left( {\cal I}_{ \mysingle }^{ \phi, \omega} [ m,\ell_{m-1} + 1,\ell_m ]  \right) \leq \opt\left( {\cal I}_{ \mysingle }^{ \phi, \omega} [ m,\ell_{m-1} + 1,\ell_m ]  \right) \ ,
    \end{equation}
    while ensuring that its profit is at least $(1-\eps)\cdot \phi$. In this context, $\rho$ stands for the ratio between the largest and smallest $\lambda^{t+}$ values within the single-cluster instance ${\cal I}_{ \mysingle }^{ \phi, \omega} [ m,\ell_{m-1} + 1,\ell_m ]$. As noted in Section~\ref{subsec:clustering}, these values differ by a factor of at most $(n/\eps)^{(1/\eps)-1}$. Hence, by plugging in $\rho = (n/\eps)^{1/\eps}$, we infer that the time required to approximate each such instance is $O(n^{ \tilde{O}( 1 / \eps^3 ) } \cdot |{\cal I}|^{ O(1) } )$.
\end{itemize}

\paragraph{The approximate formulation.} These two alterations allow us to formulate an approximate dynamic program $\tilde{F}$, defined by mimicking our exact program $F$. Specifically, for every discretized state $(m,\ell_m,\varphi_{m}) \in {\cal D}$, we have
\begin{eqnarray*}
\tilde{F}\left(m,\ell_m, \varphi_m\right)  = &\min\limits_{\ell_{m-1},{\varphi}_{m-1}} & \left\{ \tilde{F}\left(m-1,\ell_{m-1},\varphi_{m-1}\right) + \widetilde{\opt} \left( {\cal I}_{ \mysingle }^{ \phi, \omega} [ m,\ell_{m-1} + 1,\ell_m ] \right) \right\} \\
&{\mathrm{s.t.}} & \ell_{m-1} \in [\ell_m]_0 \ , \ \varphi_{m-1} \in [0, \varphi_m] \cap \Phi \\
& &\phi = \left[ {\varphi}_{m}-\left(1+\frac{\eps}{M}\right) \cdot {\varphi}_{m-1} - \delta \right]^+ \ , \\
& & \omega = \tilde{F}\left(m-1,\ell_{m-1},\varphi_{m-1}\right) \ .
\end{eqnarray*}
To gain some intuition for the definition of $\phi$, we mention that due to discretizing the profit requirement $\varphi_m$, a certain degree of inaccuracy may potentially accumulate throughout our recursive calls. As formally explained later on, choosing $\phi = [ {\varphi}_{m}-(1+\frac{\eps}{M}) \cdot {\varphi}_{m-1} - \delta ]^+$ rather than
simply $\phi = \varphi_{m} - {\varphi}_{m-1}$ is intended to offset these inaccuracies. As before, terminal states of our approximate dynamic program correspond to $m = 0$ or $\ell_{m} = -1$, in which case $\tilde{F}(m,\ell_m,0) = 0$ by definition and $\tilde{F}(m,\ell_m,\varphi_m) = \infty$ for $\varphi_m > 0$.

\paragraph{Overall running time.} Summarizing our earlier discussion, the number of states over which we compute the value function $\tilde{F}$ is only $|{\cal D}| = O(M L\cdot |\Phi|) = O(\frac{1}{\eps}\log \frac{1}{\eps} \cdot  |{\cal I}|^{ O(1) } )$. At each state, solving its recursive equation involves $O(M \cdot |{\cal D}| )$ calls to the subroutine given by Theorem~\ref{thm:main_bounded_lambda}, which approximates single-cluster instances of the inverse incremental knapsack problem. As previously noted, each such call can be executed in $O(n^{ \tilde{O}( 1 / \eps^3 ) } \cdot |{\cal I}|^{ O(1) } )$ time. All in all, our approximate dynamic programming formulation $\tilde{F}$ requires $O(n^{ \tilde{O}( 1 / \eps^3 ) } \cdot |{\cal I}|^{ O(1) } )$ time to be solved.

\paragraph{Final solution.} Lastly, it remains to explain how the various $\tilde{F}(\cdot, \cdot, \cdot)$ values we obtain can ultimately be translated back into an uncrossing-stars solution. To this end, let $\tilde{\varphi}_M$ be the maximum-possible profit $\varphi_M \in \Phi$ for which $\tilde{F}(M,L,{\varphi}_M) < \infty$. Given this quantity, consider the sequence $(0, \tilde{\ell}_0 ,\tilde{\varphi}_0), (1, \tilde{\ell}_1, \tilde{\varphi}_1), \ldots, ( M, \tilde{\ell}_M, \tilde{\varphi}_M)$ of states traversed by the dynamic program $\tilde{F}$ in reaching state $(M,\tilde{\ell}_M,\tilde{\varphi}_M)$, with the convention that $\tilde{\ell}_0 = -1$, $\tilde{\varphi}_0 = 0$, and $\tilde{\ell}_M = L$. Our solution $\bar{{\cal S}} = \langle \bar{S}_1,\ldots,\bar{S}_T \rangle $ is formed by gluing together the single-cluster solutions generated along this sequence. In other words, for each state $(m, \tilde{\ell}_m, \tilde{\varphi}_m)$, let $\tilde{\cal S}^m$ be the solution obtained by approximating the single-cluster instance ${\cal I}_{ \mysingle }^{ \tilde{\phi}_m, \tilde{\omega}_m} [ m,\tilde{\ell}_{m-1} + 1,\tilde{\ell}_m ]$, where $\tilde{\phi}_m = [ \tilde{\varphi}_{m}-(1+\frac{\eps}{M}) \cdot \tilde{\varphi}_{m-1} - \delta ]^+$ and $\tilde{\omega}_m = \tilde{F}( m-1, \tilde{\ell}_{m-1}, \tilde{\varphi}_{m-1})$. Then, at any time period $t \in {\cal C}_m$, the items introduced by $\bar{{\cal S}}$ are simply those introduced by $\tilde{\cal S}^m$, i.e., $\bar{S}_t \setminus \bar{S}_{t-1} = \tilde{S}^m_t \setminus \tilde{S}^m_{t-1}$.

\subsection{Analysis} \label{subsec:analysis-stars}

\paragraph{Feasibility.}  Prior to examining how profitable $\bar{{\cal S}}$ is, we address a more basic question, asking why this solution is indeed feasible. The next lemma argues that our construction is guaranteed to satisfy the capacity constraints of all time periods.

\begin{lemma}
$w( \bar{S}_t ) \leq W_t$ for every $t \in [T]$.
\end{lemma}
\proof{\em Proof.}
We establish the desired claim by induction over the cluster indices $m \in [M]$. For the base case of $m =1$, note that $\tilde{\cal S}_1$ is a solution to the single-cluster instance ${\cal I}_{ \mysingle }^{ \tilde{\phi}_1, \tilde{\omega}_1} [ 1,\tilde{\ell}_{0} + 1,\tilde{\ell}_1 ]$, where $\tilde{\omega}_1 = \tilde{F}(0,\tilde{\ell}_0,\tilde{\varphi}_{0}) = \tilde{F}(0,-1,0) = 0$ in view of our terminal conditions. Consequently, for every time period $t\in {\cal C}_1$, it follows that $w(\bar{S}_t) = w(\tilde{S}^1_t) \leq [W_t - \tilde{\omega}_1]^+ = W_t$, where the latter inequality holds since this period was given an adjusted capacity of $[W_t - \tilde{\omega}_1]^+$ within ${\cal I}_{ \mysingle }^{ \tilde{\phi}_1, \tilde{\omega}_1} [ 1,\tilde{\ell}_{0} + 1,\tilde{\ell}_1 ]$, as explained in Section~\ref{subsec:dp-stars}.

To prove the inductive step, with respect to a cluster index $m \geq 2$, the construction of $\bar{\cal S}$ implies that, for every time period $t \in {\cal C}_m$,
\[
\bar{S}_t \setminus \bar{S}_{\tau_{m-1}} ~~=~~ \biguplus_{q = \tau_{m-1}+1}^{t} \left(\bar{S}^{m}_{q} \setminus \bar{S}^{m}_{q-1}\right) ~~=~~ \biguplus_{q = \tau_{m-1}+1}^{t} \left(\tilde{S}^{m}_{q} \setminus \tilde{S}^{m}_{q-1}\right) ~~=~~ \tilde{S}^m_t  \ ,
\]
with the convention that $\tilde{S}^{m}_{\tau_{m-1}} = \emptyset$. Consequently,
\begin{eqnarray}
w\left(\bar{S}_t\right) & = & w\left(\bar{S}_{\tau_{m-1}}\right) + w\left(\bar{S}_t \setminus \bar{S}_{\tau_{m-1}}\right) \nonumber \\
& = & \sum_{\mu \in [m-1]} w ( \tilde{S}^{\mu}_{ \tau_{ \mu } } ) +
w(\tilde{S}^m_t) \nonumber \\
& = & \sum_{\mu \in [m-1]} \widetilde{\opt} \left( {\cal I}_{ \mysingle }^{ \tilde{\phi}_{\mu}, \tilde{\omega}_{\mu}} [ \mu,\tilde{\ell}_{\mu-1} + 1,\tilde{\ell}_{\mu} ] \right) + w(\tilde{S}^m_t) \nonumber \\
& = & \tilde{F}(m-1,\tilde{\ell}_{m-1},\tilde{\varphi}_{m-1} )  + w(\tilde{S}^m_t)  \label{ineq-w-bar} \ ,
\end{eqnarray}
where the last equality is obtained by expanding the recursive equation of our approximate dynamic program $\tilde{F}$. However, $\tilde{\omega}_{m} = \tilde{F}(m-1,\tilde{\ell}_{m-1},\tilde{\varphi}_{m-1})$ by definition, while $w(\tilde{S}^m_t) \leq [W_t - \tilde{\omega}_m]^+ = W_t - \tilde{\omega}_m$, since $\tilde{\cal S}^m$ is a solution to the single-cluster instance ${\cal I}_{ \mysingle }^{ \tilde{\phi}_m, \tilde{\omega}_m} [ m,\tilde{\ell}_{m-1} + 1,\tilde{\ell}_m ]$. By plugging these observations into~\eqref{ineq-w-bar}, we have just shown that $w(\bar{S}_t) \leq W_t$. \halmos
\endproof

\paragraph{Optimality gap.} We now turn our attention to proving that $\bar{{\cal S}}$ is profit-wise comparable to the optimal incremental knapsack solution ${\cal S}^*$, thereby concluding the proof of Theorem~\ref{thm:general}. Interestingly, a crucial part of our overall argument shows that, over the discretized state space ${\cal D}$, the value function $\tilde{F}$ provides a lower bound on the original value function $F$.

\begin{lemma} \label{lem:bound-bar-S}
$\Psi(\bar{{\cal S}}) \geq (1-7\eps)\cdot\Psi({\cal S}^*)$.
\end{lemma}
\proof{\em Proof.}
Our analysis proceeds by establishing two claims, one relating the profit $\Psi(\bar{{\cal S}})$ to the profit parameter $\tilde{\varphi}_M$, as defined in Section~\ref{subsec:adp-stars},
and the other relating between the value functions $F$ and $\tilde{F}$. Since the arguments involved are rather technical, we provide both proofs in Appendices~\ref{app:proof_clm_profit_vs_phi} and~\ref{app:proof_clm_lb-tilde-U}.

\begin{claim} \label{clm:profit_vs_phi}
$\Psi(\bar{{\cal S}}) \geq (1-2\eps) \cdot  \tilde{\varphi}_{M} - M \delta$.
\end{claim}

\begin{claim} \label{clm:lb-tilde-U}
$\tilde{F}(m,\ell_m,\varphi_m) \leq F(m,\ell_m,\varphi_m)$, for every state $(m,\ell_m,\varphi_m) \in {\cal D}$.
\end{claim}

We remind the reader that, as an intermediate step, we have shown in Corollary~\ref{cor:f_vs_opt} that $F(M, L, \varphi_M) <  \infty$ for any $\varphi_M \leq (1 - 2\eps) \cdot \objfunc( {\cal S}^* )$. In turn, Claim~\ref{clm:lb-tilde-U} implies that $\tilde{F}(M, L, \varphi_M) <  \infty$ for any $\varphi_M \in \Phi \cap [0, (1 - 2\eps) \cdot \objfunc( {\cal S}^* )]$. Therefore, since $\tilde{\varphi}_M$ is the maximum-possible profit $\varphi_M \in \Phi$ for which $\tilde{F}(M,L,{\varphi}_M) < \infty$, we clearly have
\[ \tilde{\varphi}_M ~~\geq~~ \max \left\{ \Phi \cap [0, (1 - 2\eps) \cdot \objfunc( {\cal S}^* )] \right\} ~~\geq~~ \frac{ (1 - 2\eps) \cdot \objfunc( {\cal S}^* ) }{ 1+\frac{\eps}{M} } - \delta ~~\geq~~ (1 - 3\eps) \cdot \objfunc( {\cal S}^* ) - \delta \ , \]
where the middle inequality follows from the definition of $\Phi$ in~\eqref{eqn:def_phi}. Plugging this relation back into Claim~\ref{clm:profit_vs_phi}, we derive the desired bound on $\Psi(\bar{{\cal S}})$ by observing that
\begin{eqnarray*}
\Psi(\bar{{\cal S}}) & \geq & (1-2\eps) \cdot  \tilde{\varphi}_{M} - M \delta \\
& \geq & (1-5\eps) \cdot \objfunc( {\cal S}^* ) - (M + 1) \cdot \delta \\
& \geq & (1-5\eps) \cdot \objfunc( {\cal S}^* ) - 2 \eps \cdot \lambda_T p_{ \max } \\
& \geq & (1-7\eps) \cdot \objfunc( {\cal S}^* ) \ .
\end{eqnarray*}
Here, the third inequality holds since $\delta = \frac{\eps}{M} \cdot\lambda_T\cdot p_{ \max }$, and the fourth inequality is obtained by noting that $\lambda_T  p_{ \max } \leq \Psi({\cal S}^*)$, since an objective value of $\lambda_T  p_{ \max }$ can trivially be attained for the incremental knapsack problem and since ${\cal S}^*$ is an optimal solution. \halmos
\endproof

\section{Concluding Remarks}

\paragraph{Subsequent work.} Following a preliminary version of the current paper, \cite{FaenzaSZ20} have investigated the generalized incremental knapsack problem, in which item profits have a general dependency on their insertion time. In other words, letting $p_{it}$ be the marginal profit contribution due to introducing item $i$ at time period $t$, the quantities $\{ p_{it} \}_{t \in [T]}$ are allowed to be completely unrelated. In our setting, these contributions take a simpler product form, $p_{it} = p_i \cdot \lambda^{t+}$. Quite surprisingly, \cite{FaenzaSZ20} proved that generalized incremental knapsack can be approximated within factor $\frac{ 1 }{ 2 } - \eps$ in polynomial time. Additionally, the authors have proposed a quasi-PTAS, computing $(1-\eps)$-approximate solutions in slightly super-polynomial time; specifically, this approach includes an exponential dependency on $O(\log^{O(1)} |{\cal I}|)$, where $|{\cal I}|$ stands for the input size in binary representation. Even though the latter bounds are weaker in absolute terms than the polynomial-time approximation scheme we have developed here, it is worth keeping in mind that these results are incomparable since, as the name implies, the generalized incremental knapsack problem is a more general computational setting. From a technical standpoint, the methods involved are very different in nature, and we are not aware of any way to leverage our findings for the purpose of devising a PTAS for generalized incremental knapsack or vice versa.

\ACKNOWLEDGMENT{We thank Jacob Feldman (WUSTL) for his feedback on an earlier version of this paper.}

\bibliographystyle{informs2014}
\bibliography{BIB-INC-KNAPSACK}

\begin{APPENDICES}
\section{Additional Proofs}

\subsection{Proof of Lemma~\ref{lem:properties_up_rounding}} \label{subsec:proof_lem_properties_up_rounding}

\paragraph{Proof of item~\ref{lem:rounding_monotone}.} Given a pair of utilization vectors, $N \leq N^+$, we show that $N^{(1)} \leq N^{+(1)}$ by considering two cases, depending on whether a given profit class is light or heavy for $N$.

\noindent {\em Case 1: $\ell \in \mylight(N)$}. This is the easy scenario, since
\[ N^{(1)}_{ \ell } ~~=~~ N_{ \ell } ~~\leq~~ N^+_{ \ell } ~~\leq~~ N^{+(1)}_{ \ell } \ , \]
where the first equality is simply the definition of $N^{(1)}_{ \ell }$ for light classes and the last inequality is due to having $N^+ \leq N^{+(1)}$, following the up-rounding operation.

\noindent {\em Case 2: $\ell \in \myheavy(N)$}. We begin by noting that, since $N \leq N^+$, it follows that $\ell \in \myheavy(N^+)$ as well. Consequently, by definition of $N^{(1)}_{ \ell }$ and $N^{+(1)}_{ \ell }$, we know that these values are precisely:
\begin{eqnarray*}
N^{(1)}_{ \ell } & = & \max \left\{ k \in \left[ \frac{ 1 }{ \eps } + 1, | {\cal P}_{ \ell } | \right] : w \left( {\cal P}_{ \ell } \left[ \frac{ 1 }{ \eps } + 1, k \right] \right) \leq {\cal E}_{ \ell } \right\} \ , \\
N^{+(1)}_{ \ell } & = & \max \left\{ k \in \left[ \frac{ 1 }{ \eps } + 1, | {\cal P}_{ \ell } | \right] : w \left( {\cal P}_{ \ell } \left[ \frac{ 1 }{ \eps } + 1, k \right] \right) \leq {\cal E}_{ \ell }^+ \right\} \ .
\end{eqnarray*}
While these expressions are nearly identical, the estimate we make use of in the $N^{(1)}_{ \ell }$-related expression is ${\cal E}_{ \ell } = \mu_{ \ell } \cdot \mypower_2 [ \frac{ \eps }{ |{\cal L}| } \cdot W_{ \myheavy }( N ) ]$, where $\mu_{ \ell } \geq 1$ is the unique integer for which
\[ \left( \mu_{ \ell } - 1 \right) \cdot \mypower_2 \left[ \frac{ \eps }{ |{\cal L}| } \cdot W_{ \myheavy }( N ) \right] ~~<~~ w \left( {\cal P}_{ \ell } \left[ \frac{ 1 }{ \eps } + 1, N_{\ell} \right] \right) ~~\leq~~ \mu_{ \ell } \cdot \mypower_2 \left[ \frac{ \eps }{ |{\cal L}| } \cdot W_{ \myheavy }( N ) \right] \ . \]
On the other hand, in the $N^{+(1)}_{ \ell }$-related expression, our estimate is ${\cal E}_{ \ell }^+ = \mu_{ \ell }^+ \cdot \mypower_2 [ \frac{ \eps }{ |{\cal L}| } \cdot W_{ \myheavy }( N^+ ) ]$, where $\mu_{ \ell }^+  \geq 1$ is the unique integer for which
\[ \left( \mu_{ \ell }^+ - 1 \right) \cdot \mypower_2 \left[ \frac{ \eps }{ |{\cal L}| } \cdot W_{ \myheavy }( N^+ ) \right] ~~<~~ w \left( {\cal P}_{ \ell } \left[ \frac{ 1 }{ \eps } + 1, N_{\ell}^+ \right] \right) ~~\leq~~ \mu_{ \ell }^+ \cdot \mypower_2 \left[ \frac{ \eps }{ |{\cal L}| } \cdot W_{ \myheavy }( N^+ ) \right] \ . \]
Therefore, to show that $N^{(1)}_{ \ell } \leq N^{+(1)}_{ \ell }$, it remains to prove that ${\cal E}_{ \ell } \leq {\cal E}_{ \ell }^+$. For this purpose, note that
\begin{eqnarray*}
\left( \mu_{ \ell } - 1 \right) \cdot \mypower_2 \left[ \frac{ \eps }{ |{\cal L}| } \cdot W_{ \myheavy }( N ) \right] & < & w \left( {\cal P}_{ \ell } \left[ \frac{ 1 }{ \eps } + 1, N_{\ell} \right] \right) \\
& \leq & w \left( {\cal P}_{ \ell } \left[ \frac{ 1 }{ \eps } + 1, N_{\ell}^+ \right] \right) \\
& \leq & \mu_{ \ell }^+ \cdot \mypower_2 \left[ \frac{ \eps }{ |{\cal L}| } \cdot W_{ \myheavy }( N^+ ) \right] \ ,
\end{eqnarray*}
where the second inequality holds since $N \leq N^+$. By rearranging the bound we have just attained, it follows that $\mu_{ \ell } - 1 < \mu_{ \ell }^+ \cdot \frac{ \mypower_2 [ \frac{ \eps }{ |{\cal L}| } \cdot W_{ \myheavy }( N^+ ) ] }{ \mypower_2 [ \frac{ \eps }{ |{\cal L}| } \cdot W_{ \myheavy }( N ) ] }$; for this transition, note that $\mypower_2 [ \frac{ \eps }{ |{\cal L}| } \cdot W_{ \myheavy }( N ) ] > 0$, as $W_{ \myheavy }( N ) \geq w ( {\cal P}_{ \ell } [ \frac{ 1 }{ \eps } + 1, N_{\ell} ] ) \geq \min_{i \in {\cal P}_{ \ell }} w_i > 0$, by our initial assumption that all item weights are strictly positive (see Section~\ref{subsec:model_description}). However, since $W_{ \myheavy }( N^+ ) \geq W_{ \myheavy }( N )$, we know that $\frac{ \mypower_2 [ \frac{ \eps }{ |{\cal L}| } \cdot W_{ \myheavy }( N^+ ) ] }{ \mypower_2 [ \frac{ \eps }{ |{\cal L}| } \cdot W_{ \myheavy }( N ) ] }$ is an integer. Moreover, since $\mu_{ \ell }$ and $\mu_{ \ell }^+$ are integers as well, we necessary have $\mu_{ \ell } \leq \mu_{ \ell }^+ \cdot \frac{ \mypower_2 [ \frac{ \eps }{ L } \cdot W_{ \myheavy }( N^+ ) ] }{ \mypower_2 [ \frac{ \eps }{ L } \cdot W_{ \myheavy }( N ) ] }$. This inequality is precisely ${\cal E}_{ \ell } \leq {\cal E}_{ \ell }^+$, by definition of these two estimates.

\paragraph{Proof of item~\ref{lem:rounding_weight}.} To obtain the desired upper bound on $W_{ \myheavy }( N^{(1)} )$, note that
\begin{eqnarray*}
W_{ \myheavy } \left( N^{(1)} \right) & = & \sum_{ \ell \in \myheavy( N^{(1)} ) } w \left( {\cal P}_{ \ell } \left[ \frac{ 1 }{ \eps } + 1, N^{(1)}_{\ell} \right] \right) \\
& \leq & \sum_{ \ell \in \myheavy( N ) } {\cal E}_{ \ell } \\
& \leq & \sum_{ \ell \in \myheavy( N ) } \left( w \left( {\cal P}_{ \ell } \left[ \frac{ 1 }{ \eps } + 1, N_{\ell} \right] \right) + \mypower_2 \left[ \frac{ \eps }{ |{\cal L}| } \cdot W_{ \myheavy }( N ) \right] \right) \\
& \leq & W_{ \myheavy }( N ) + 2\eps \cdot \frac{ | \myheavy( N ) | }{ |{\cal L}| } \cdot W_{ \myheavy }( N ) \\
& \leq & (1 + 2\eps) \cdot W_{ \myheavy }( N ) \ .
\end{eqnarray*}
Here, the first inequality holds since $\myheavy( N^{(1)} ) = \myheavy( N )$ and since $w ( {\cal P}_{ \ell } [ \frac{ 1 }{ \eps } + 1, N^{(1)}_{ \ell } ] ) \leq {\cal E}_{ \ell }$, by definition of $N^{(1)}_{ \ell }$. The second inequality is obtained by recalling that ${\cal E}_{ \ell } - \mypower_2 [ \frac{ \eps }{ |{\cal L}| } \cdot W_{ \myheavy }( N ) ] < w ( {\cal P}_{ \ell } [ \frac{ 1 }{ \eps } + 1, N_{\ell} ] )$.

\subsection{Proof of Lemma~\ref{lem:properties_truncation}} \label{subsec:proof_lem_properties_truncation}

\paragraph{Proof of item~\ref{lem:truncation_monotone}.} Given a pair of utilization vectors, $N \leq N^+$, we have already shown in Lemma~\ref{lem:properties_up_rounding}\ref{lem:rounding_monotone} that $N^{(1)} \leq N^{+(1)}$. In order to further argue that $N^{(2)} \leq N^{+(2)}$, our proof proceeds by considering four cases, depending on whether a given profit class is light or heavy for $N^{(1)}$ and $N^{+(1)}$.
\begin{enumerate}
\item {\em $\ell \in \mylight(N^{(1)})$ and $\ell \in \mylight(N^{+(1)})$}. In this scenario,
    \[ N^{(2)}_{ \ell } ~~=~~ N^{(1)}_{ \ell } ~~\leq~~ N^{+(1)}_{ \ell } ~~=~~ N^{+(2)}_{ \ell } \ , \]
    where the first and last equalities follow from how the truncation operation is defined for light profit classes, and the middle inequality holds since $N^{(1)} \leq N^{+(1)}$.

\item {\em $\ell \in \mylight(N^{(1)})$ and $\ell \in \myheavy(N^{+(1)})$}. As explained in Section~\ref{subsec:truncation}, the truncation operation is class-preserving, meaning in particular that $\ell \in \mylight(N^{(2)})$ and $\ell \in \myheavy(N^{+(2)})$. Therefore, $N^{(2)}_{ \ell } \leq \frac{ 1 }{ \eps } < N^{+(2)}_{ \ell }$.

\item {\em $\ell \in \myheavy(N^{(1)})$ and $\ell \in \mylight(N^{+(1)})$}. This scenario is not possible, since $N^{(1)} \leq N^{+(1)}$.

\item {\em $\ell \in \myheavy(N^{(1)})$ and $\ell \in \myheavy(N^{+(1)})$}. In this case, we begin by recalling that $N^{(2)}_{ \ell } = N^{(1)}_{ \ell } - \lceil 2\eps \cdot \Delta_{ \ell } \rceil$ and $N^{+(2)}_{ \ell } = N^{+(1)}_{ \ell } - \lceil 2\eps \cdot \Delta_{ \ell }^+ \rceil$, where $\Delta_{ \ell } = N^{(1)}_{ \ell } - \frac{ 1 }{ \eps }$ and $\Delta_{ \ell }^+ = N^{+(1)}_{ \ell } - \frac{ 1 }{ \eps }$, noting that $\Delta_{ \ell } \leq \Delta_{ \ell }^+$ as $N^{(1)} \leq N^{+(1)}$. Now, when $\Delta_{ \ell } = \Delta_{ \ell }^+$, we immediately get $N^{(2)}_{ \ell } \leq N^{+(2)}_{ \ell }$. In the complementary case, where $\Delta_{ \ell } \leq \Delta_{ \ell }^+ - 1$, note that
    \begin{eqnarray*}
    N^{+(2)}_{ \ell } & = & N^{+(1)}_{ \ell } - \lceil 2\eps \cdot \Delta_{ \ell }^+ \rceil \\
    & = & \frac{ 1 }{ \eps } + \Delta_{ \ell }^+ - \lceil 2\eps \cdot \Delta_{ \ell }^+ \rceil \\
    & \geq & \frac{ 1 }{ \eps } + (1 - 2\eps) \cdot \Delta_{ \ell }^+ - 1 \\
    & \geq & \frac{ 1 }{ \eps } + (1 - 2\eps) \cdot \left( \Delta_{ \ell } + 1 \right) - 1 \\
    & \geq & \frac{ 1 }{ \eps } + \Delta_{ \ell } - \lceil 2\eps \cdot \Delta_{ \ell } \rceil - 2 \eps \\
    & = & N^{(2)}_{ \ell } - 2 \eps \ .
    \end{eqnarray*}
    Thus, since $\eps < 1/2$, we indeed have $N^{(2)}_{ \ell } \leq N^{+(2)}_{ \ell }$.
\end{enumerate}

\paragraph{Proof of item~\ref{lem:truncation_weight}.} To prove that $w( N^{(2)} ) \leq w( N )$, we first express these quantities as:
\begin{eqnarray*}
w \left( N^{(2)} \right) & = & \sum_{ \ell \in \mylight(  N^{(2)} ) } w \left( {\cal P}_{ \ell } \left[ 1, N^{(2)}_{\ell} \right] \right) + \sum_{ \ell \in \myheavy(  N^{(2)} ) } w \left( {\cal P}_{ \ell } \left[ 1, \frac{ 1 }{ \eps } \right] \right) \\
& & \mbox{} + \underbrace{ \sum_{ \ell \in \myheavy( N^{(2)} ) } w \left( {\cal P}_{ \ell } \left[ \frac{ 1 }{ \eps } + 1, N^{(2)}_{\ell} \right] \right) }_{ W_{ \myheavy }( N^{(2)} ) } \ , \\
w \left( N \right) & = & \sum_{ \ell \in \mylight(  N ) } w \left( {\cal P}_{ \ell } \left[ 1, N_{\ell} \right] \right) + \sum_{ \ell \in \myheavy(  N ) } w \left( {\cal P}_{ \ell } \left[ 1, \frac{ 1 }{ \eps } \right] \right) \\
& & \mbox{} + \underbrace{ \sum_{ \ell \in \myheavy( N ) } w \left( {\cal P}_{ \ell } \left[ \frac{ 1 }{ \eps } + 1, N_{\ell} \right] \right) }_{ W_{ \myheavy }( N ) } \ .
\end{eqnarray*}
As explained in Section~\ref{subsec:truncation}, we know that the light and heavy classes of $N^{(2)}$ and $N$ are identical, and therefore, the first two summands in each of the above expressions are identical, noting that $N^{(2)}_{\ell} = N_{\ell}$ for every $\ell \in \mylight( N^{(2)} ) = \mylight( N )$. Consequently, it suffices to show that $W_{ \myheavy }( N^{(2)} ) \leq W_{ \myheavy }( N )$.

To this end, let us consider some class $\ell \in \myheavy( N )$. Recall that $N^{(2)}_{ \ell } = N^{(1)}_{ \ell } - \lceil 2\eps \cdot \Delta_{ \ell } \rceil$ and that the internal order between the items in ${\cal P}_{\ell}$ is increasing by weight. Therefore, by eliminating the $\lceil 2\eps \cdot \Delta_{ \ell } \rceil$ heaviest items picked by $N^{(1)}_{ \ell }$ from this class, we have deleted a fraction of at least $\frac{ \lceil 2\eps \cdot \Delta_{ \ell } \rceil }{ \Delta_{ \ell } } \geq 2 \eps$ of the total weight $w ( {\cal P}_{ \ell } [ \frac{ 1 }{ \eps } + 1, N^{(1)}_{\ell} ] )$. In other words, $w ( {\cal P}_{ \ell } [ \frac{ 1 }{ \eps } + 1, N^{(2)}_{\ell} ] ) \leq (1 - 2\eps) \cdot w ( {\cal P}_{ \ell } [ \frac{ 1 }{ \eps } + 1, N^{(1)}_{\ell} ] )$. Summing this inequality over all heavy classes, it follows that
\[ W_{ \myheavy } \left( N^{(2)} \right) ~~\leq~~ (1 - 2\eps) \cdot W_{ \myheavy } \left( N^{(1)} \right) ~~\leq~~ (1 - 2\eps) \cdot (1 + 2\eps) \cdot W_{ \myheavy } \left( N \right) ~~\leq~~ W_{ \myheavy } \left( N \right) \ , \]
where the second inequality follows from Lemma~\ref{lem:properties_up_rounding}\ref{lem:rounding_weight}.

\paragraph{Proof of item~\ref{lem:truncation_picks_heavy}.} To prove that $N^{(2)}_{ \ell } \geq (1 - 2\eps ) \cdot N_{ \ell }$ for every $\ell \in \myheavy( N )$, we initially recall that $N \leq N^{(1)}$, as shown in Section~\ref{subsec:up_rounding}. Hence, it suffices to argue that $N^{(2)}_{ \ell } \geq (1 - 2\eps ) \cdot N^{(1)}_{ \ell }$. To this end, since  $N^{(2)}_{ \ell } = N^{(1)}_{ \ell } - \lceil 2\eps \cdot \Delta_{ \ell } \rceil$ and $\Delta_{ \ell } = N^{(1)}_{ \ell } - \frac{ 1 }{ \eps }$, the latter inequality follows by observing that
\[ \lceil 2\eps \cdot \Delta_{ \ell } \rceil ~~\leq~~ 2\eps \cdot \Delta_{ \ell } + 1 ~~=~~ 2\eps \cdot \left( N^{(1)}_{ \ell } - \frac{ 1 }{ \eps } \right) + 1 ~~\leq~~ 2\eps \cdot N^{(1)}_{ \ell } \ . \]

\subsection{Proof of Claim~\ref{clm:ub_sum_mu}} \label{app:proof_clm_ub_sum_mu}

We prove the desired bound by contradiction. To this end, suppose that we actually have $\sum_{ \ell \in \myheavy( N ) } \mu_{ \ell } > \frac{ 3 }{ 2 } \cdot \frac{ |{\cal L}| }{ \eps }$. Then,
\begin{eqnarray*}
& & \sum_{ \ell \in \myheavy( N ) } \left( \mu_{ \ell } - 1 \right) \cdot \mypower_2 \left[ \frac{ \eps }{ |{\cal L}| } \cdot W_{ \myheavy }( N ) \right] \\
& & \qquad \qquad \geq \frac{ \eps }{ |{\cal L}| } \cdot W_{ \myheavy }( N ) \cdot \sum_{ \ell \in \myheavy( N ) } \mu_{ \ell } - 2 \eps \cdot \frac{ | \myheavy( N ) | }{ |{\cal L}| } \cdot W_{ \myheavy }( N ) \\
& & \qquad \qquad \geq \Bigg( \frac{ \eps }{ |{\cal L}| } \cdot \sum_{ \ell \in \myheavy( N ) } \mu_{ \ell } - \frac{ 1 }{ 2 } \Bigg) \cdot W_{ \myheavy }( N ) \\
& & \qquad \qquad > W_{ \myheavy }( N ) \\
& & \qquad \qquad = \sum_{ \ell \in \myheavy( N ) } w \left( {\cal P}_{ \ell } \left[ \frac{ 1 }{ \eps } + 1, N_{\ell} \right] \right) \ ,
\end{eqnarray*}
where the second and third inequalities hold since $\eps \leq 1/4$ and $\sum_{ \ell \in \myheavy( N ) } \mu_{ \ell } > \frac{ 3 }{ 2 } \cdot \frac{ |{\cal L}| }{ \eps }$, respectively. Therefore, there is at least one index $\ell \in \myheavy( N )$ for which $( \mu_{ \ell } - 1 ) \cdot \mypower_2 [ \frac{ \eps }{ |{\cal L}| } \cdot W_{ \myheavy }( N ) ] > w ( {\cal P}_{ \ell } [ \frac{ 1 }{ \eps } + 1, N_{\ell} ] )$, contradicting the definition of $\mu_{ \ell }$ in equation~\eqref{eqn:def_mu}.

\subsection{Proof of Claim~\ref{clm:profit_vs_phi}} \label{app:proof_clm_profit_vs_phi}

To obtain the desired bound, we observe that the total profit attained by $\bar{\cal S}$ can be decomposed as follows:
\begin{eqnarray}
\Psi\left(\bar{{\cal S}}\right) &=&	 \sum_{m \in [M]} \Psi( \tilde{\cal S}^m) \label{ineq:def}\\
&\geq & \left(1-\eps\right) \cdot \sum_{m \in [M]} \tilde{\phi}_m \label{ineq:f1}\\
& \geq & \left(1-\eps\right) \cdot \sum_{m \in [M]} \left( \tilde{\varphi}_{m}-\left(1+\frac{\eps}{M}\right) \cdot \tilde{\varphi}_{m-1} - \delta\right)\label{ineq:f1.5}\\
& \geq & \left(1-\eps\right) \cdot  \tilde{\varphi}_{M} -  \frac{\eps}{M} \cdot \sum_{m \in [M-1]} \tilde{\varphi}_{m}- M \delta \label{ineq:f2}\\
& \geq & \left(1-2\eps\right) \cdot  \tilde{\varphi}_{M} - M \delta \label{ineq:f3} \ .
\end{eqnarray}
Equality~\eqref{ineq:def} holds since, by definition of $\bar{\cal S}$, each item introduced by $\bar{\cal S}$ appears in exactly one of the single-cluster solutions $\tilde{\cal S}^1, \ldots, \tilde{\cal S}^M$, where it is introduced at precisely the same time period. To derive inequality~\eqref{ineq:f1}, note that since $\tilde{\cal S}^m$ is obtained by applying the approximation scheme proposed in Theorem~\ref{thm:main_bounded_lambda} with respect to the single-cluster instance ${\cal I}_{ \mysingle }^{ \tilde{\phi}_m, \tilde{\omega}_m} [ m,\tilde{\ell}_{m-1} + 1,\tilde{\ell}_m ]$, we are guaranteed to collect a profit of $\Psi( \tilde{\cal S}^m) \geq (1 - \eps) \cdot \tilde{\phi}_m$. Inequality~\eqref{ineq:f1.5} holds since $\tilde{\phi}_m = [ \tilde{\varphi}_{m}-(1+\frac{\eps}{M}) \cdot \tilde{\varphi}_{m-1} - \delta ]^+$. Inequality~\eqref{ineq:f2} proceeds by noting that $\tilde{\varphi}_0 = 0$. Inequality~\eqref{ineq:f3} is obtained by observing that $\tilde{\varphi}_1 \leq \cdots \leq \tilde{\varphi}_M$, as the dynamic program $\tilde{F}$ is forced to pick each $\tilde{\varphi}_{m-1}$ out of $[0, \tilde{\varphi}_m] \cap \Phi$.

\subsection{Proof of Claim~\ref{clm:lb-tilde-U}} \label{app:proof_clm_lb-tilde-U}

We prove the claim by induction over the discretized state space ${\cal D}$, following the order by which states are considered in the approximate program $\tilde{F}$. The base case corresponds to terminal states $(m,\ell_m,\varphi_m)$, for which $m = 0$ or $\ell_m = -1$. Here, we have $\tilde{F}(m,\ell_m,\varphi_m) = F(m,\ell_m,\varphi_m)$, according to the terminal conditions of $F$ and $\tilde{F}$. To establish the induction step, consider a state $(m,\ell_m, \varphi_m) \in {\cal D}$ with $m \geq 1$ and $\ell_m \geq 0$. Let $(\ell^*_{m-1}, {\varphi}^*_{m-1})$ be an optimal solution to the minimization problem given by our recursive equation for $F(m,\ell_m,\varphi_m)$. Next, we define $\varphi^{\Phi}_{m-1}$ as the result of rounding $\varphi_{m-1}^*$ down to the nearest value in $\Phi$, which implies that
\begin{eqnarray} \label{eq:accuracy-var}
\varphi^{\Phi}_{m-1} ~~\leq~~ {\varphi}^*_{m-1} ~~\leq~~ \left(1+\frac{\eps}{M}\right)\cdot \varphi^{\Phi}_{m-1} + \delta \ ,	
\end{eqnarray}
given the definition of $\Phi$ in~\eqref{eqn:def_phi}. Lastly, we define the auxiliary notation $\phi^{\Phi} = [{\varphi}_{m}-(1+\frac{\eps}{M})\cdot{\varphi}^{\Phi}_{m-1} - \delta]^+$ and $\omega^{\Phi} =  \tilde{F} (m-1, \ell^*_{m-1}, \varphi^{\Phi}_{m-1})$.
With these definitions, our recursive equation for $\tilde{F}$ ensures that
\begin{eqnarray}
\tilde{F}(m,\ell_m, \varphi_m) & \leq & \tilde{F} \left(m-1,\ell_{m-1}^*, \varphi^{\Phi}_{m-1} \right) + \widetilde{\opt} \left( {\cal I}_{ \mysingle }^{ \phi^{\Phi}, \omega^{\Phi}} [ m,\ell_{m-1}^* + 1,\ell_m ] \right) \nonumber \\
& \leq & F \left(m-1,\ell_{m-1}^*, \varphi^{\Phi}_{m-1} \right) + \widetilde{\opt} \left( {\cal I}_{ \mysingle }^{ \phi^{\Phi}, \omega^{\Phi}} [ m,\ell_{m-1}^* + 1,\ell_m ] \right) \nonumber \\
& \leq & F \left(m-1,\ell_{m-1}^*, \varphi^{\Phi}_{m-1} \right) + \opt \left( {\cal I}_{ \mysingle }^{ \phi^{\Phi}, \omega^{\Phi}} [ m,\ell_{m-1}^* + 1,\ell_m ] \right) \ . \label{eqn:UB_tilde_f}
\end{eqnarray}
Here, the second inequality proceeds from the induction hypothesis, and the third inequality employs the super-optimality property of the single-cluster solution we obtain for ${\cal I}_{ \mysingle }^{ \phi^{\Phi}, \omega^{\Phi}} [ m,\ell_{m-1}^* + 1,\ell_m ]$, as accounted for in~\eqref{eqn:opt_vs_tildeopt}.

On the other hand, letting $\phi^* = \varphi_m- {\varphi}^*_{m-1} $ and $\omega^* = {F}(m-1,\ell^*_{m-1},{\varphi}^*_{m-1})$, the optimality of $(\ell^*_{m-1}, {\varphi}^*_{m-1})$ for the minimization problem that defines $F(m,\ell_m,\varphi_m)$ implies
\begin{eqnarray}
F(m,\ell_m, \varphi_m) & = & F \left(m-1,\ell_{m-1}^*, \varphi^*_{m-1} \right) + \opt \left( {\cal I}_{ \mysingle }^{ \phi^*, \omega^*} [ m,\ell_{m-1}^* + 1,\ell_m ] \right) \nonumber \\
& \geq & F \left(m-1,\ell_{m-1}^*, \varphi^{\Phi}_{m-1} \right) + \opt \left( {\cal I}_{ \mysingle }^{ \phi^{\Phi}, \omega^{\Phi}} [ m,\ell_{m-1}^* + 1,\ell_m ] \right) \ . \label{eqn:LB_f}
\end{eqnarray}
The inequality above holds since, as we explain below, $\omega^{\Phi} \leq \omega^*$ and $\phi^{\Phi} \leq \phi^*$. Therefore, any feasible solution to ${\cal I}_{ \mysingle }^{ \phi^*, \omega^*} [ m,\ell_{m-1}^* + 1,\ell_m ]$ is also feasible to ${\cal I}_{ \mysingle }^{ \phi^{\Phi}, \omega^{\Phi}} [ m,\ell_{m-1}^* + 1,\ell_m ]$, and hence, $\opt ( {\cal I}_{ \mysingle }^{ \phi^{\Phi}, \omega^{\Phi}} [ m,\ell_{m-1}^* + 1,\ell_m ] ) \leq \opt ( {\cal I}_{ \mysingle }^{ \phi^*, \omega^*} [ m,\ell_{m-1}^* + 1,\ell_m ] )$. Noting that the right hand sides of~\eqref{eqn:UB_tilde_f} and~\eqref{eqn:LB_f} are identical, we conclude that $\tilde{F}(m,\ell_m, \varphi_m) \leq F(m,\ell_m, \varphi_m)$, as required.

Finally, to see why $\omega^{\Phi} \leq \omega^*$ as argued above, note that
\[ \omega^{\Phi} ~~=~~ \tilde{F} (m-1, \ell^*_{m-1}, \varphi^{\Phi}_{m-1}) ~~\leq~~ F (m-1, \ell^*_{m-1}, \varphi^{\Phi}_{m-1}) ~~\leq~~ F (m-1, \ell^*_{m-1}, \varphi^*_{m-1}) ~~=~~ \omega^* \ , \]
where the first inequality follows from the induction hypothesis, and the second inequality holds since $\varphi^{\Phi}_{m-1} \leq {\varphi}^*_{m-1}$ by inequality~\eqref{eq:accuracy-var}. Similarly, in order to show that $\phi^{\Phi} \leq \phi^*$, the important observation is that
\[ \phi^{\Phi} ~~=~~ \left[ {\varphi}_{m}- \left(1+\frac{\eps}{M} \right)\cdot{\varphi}^{\Phi}_{m-1} - \delta \right]^+ ~~\leq~~ \left[ {\varphi}_{m}- {\varphi}^*_{m-1} \right]^+ ~~=~~ {\varphi}_{m}- {\varphi}^*_{m-1} ~~=~~ \phi^* \ , \]
where the inequality above holds since ${\varphi}^*_{m-1} \leq (1+\frac{\eps}{M}) \cdot \varphi^{\Phi}_{m-1} + \delta$, by inequality~\eqref{eq:accuracy-var}. 
\end{APPENDICES}

\end{document}